\documentclass[prl,aps,twocolumn,a4paper,floatfix,showpacs,nofootinbib]{revtex4-1}
\usepackage[utf8]{inputenc}  
\usepackage[T1]{fontenc}
\usepackage{graphicx,psfrag}
\usepackage{mathrsfs}
\usepackage{amsmath,amsfonts,amssymb,pifont,gensymb}
\usepackage{multirow,enumerate}
\usepackage{comment,hyperref}
\usepackage{color}
\usepackage{acronym}
\usepackage{xspace}
\usepackage[normalem]{ulem}
\usepackage{mathtools}
\usepackage{subfigure}
\usepackage{makecell}
\usepackage{appendix}
\usepackage{color}

\definecolor{darkblue}{rgb}{0.0,0.0,0.6}

\begin{document}
\title{Normalizing flows as an avenue to study overlapping gravitational wave signals}

\author{Jurriaan Langendorff$^{1}$}\email[]{j.w.langendorff@uu.nl}
\author{Alex Kolmus$^{2}$}\email[]{alex.kolmus@ru.nl}
\author{Justin Janquart$^{1, 3}$}
\author{Chris Van Den Broeck$^{1, 3}$}

\affiliation{${}^1$ Institute for Gravitational and Subatomic Physics (GRASP), Department of Physics, Utrecht University, Princetonplein 1, 3584 CC Utrecht, The Netherlands}
\affiliation{${}^2$ Institute for Computing and Information Sciences (ICIS), Radboud University Nijmegen, Toernooiveld 212, 6525 EC Nijmegen, The Netherlands}
\affiliation{${}^3$ Nikhef, Science Park 105, 1098 XG Amsterdam, The Netherlands}

\date{\today}

\begin{abstract}
    Due to its speed after training, machine learning is often envisaged as a solution to a manifold of the issues faced in gravitational-wave astronomy. Demonstrations have been given for various applications in gravitational-wave data analysis. In this work, we focus on a challenging problem faced by third-generation detectors: parameter inference for overlapping signals. Due to the high detection rate and increased duration of the signals, they will start to overlap, possibly making traditional parameter inference techniques difficult to use. Here, we show a proof-of-concept application of normalizing flows to perform parameter estimation on overlapped binary black hole systems. 
\end{abstract}

\maketitle

\section{Introduction}
\label{sec:intro}
Over the last few years, the improved sensitivity of the LIGO~\cite{Aasi_2015} and Virgo~\cite{TheVirgo:2014hva} detectors has made the detection of gravitational waves (GWs) originating from compact binary coalescences (CBCs) more and more common, with over 90 detections reported after the third observation run~\cite{LIGOScientific:2021djp}. This new information channel has had major impact in fundamental physics~\cite{LIGOScientific:2021sio}, astrophysics~\cite{LIGOScientific:2021psn}, and cosmology~\cite{LIGOScientific:2021aug}. Soon, the upgrade of the current detectors and the addition of KAGRA~\cite{Somiya:2011np, Aso:2013eba, Akutsu:2018axf, Akutsu:2020his, KAGRA:2020tym} and LIGO India~\cite{LigoIndia} to the network of ground-based interferometers will lead to even more detections. In addition, the passage from second-generation (2G) to third-generation (3G) detectors (Einstein Telescope (ET)~\cite{Punturo_2010, Hild:2010id} and Cosmic Explorer (CE)~\cite{Reitze:2019iox, Abbott_2017, PhysRevLett.118.151105}) will lead to an important increase in the number of observed CBCs. These detectors are also projected to have a reduced lower frequency cutoff~\cite{Sathyaprakash:2012jk}, leading to a longer duration for the signals. Therefore, CBC signals will likely overlap in 3G detectors~\cite{Regimbau:2009rk, Samajdar:2021egv, Pizzati:2021apa, Relton:2021cax, Himemoto:2021ukb}.

It has been established that analyzing one of the overlapping signals without accounting for the presence of the other can lead to biases in the recovered posteriors, especially when the merger times of the two events are close~\cite{Samajdar:2021egv, Pizzati:2021apa, Relton:2021cax, Himemoto:2021ukb, Antonelli:2021vwg}. These could impact any direct science case for CBCs (e.g. tests of general relativity~\cite{Wu:2022pyg}), but also indirectly related ones such as the hunt for primordial black holes, where subtraction of the foreground sources is required~\cite{Sachdev2020:pol, Sharma:2022kds, Biscoveanu:2020ste, Zhou:2022nmt, Zhou:2022otw, Reali:2022aps}. 
In Ref.~\cite{Janquart:2022nyz}, the authors demonstrate on two overlapped binary black holes (BBHs) how using adapted Bayesian inference can help reduce the biases. They use two methods: (i) hierarchical subtraction, where one analyzes the dominant signal first and subtracts it before analyzing the second event, and (ii) joint parameter estimation, where the two signals are analyzed jointly. They show that the second method is less prone to biases, but computationally heavier. Their analysis also suffers some instabilities, and further upgrades are needed for it to be entirely reliable. An issue also mentioned in this work is the computational time. Indeed, more than $10^5$ CBC mergers are expected in the 3G era~\cite{Samajdar:2021egv}. So having analyses taking several weeks to complete is not a realistic alternative. 

While it is true that traditional methods can be sped-up~\cite{Zackay:2018qdy, Dai:2018dca, Leslie:2021ssu, Morisaki:2021ngj}, or that quantum computing~\cite{Gao:2021rxg} could potentially be used in the future, the development of frameworks capable of doing complete analyses in very short timescales is crucial for the development of 3G detectors. Therefore, in this work, we propose the first step in that direction, showing how overlapping BBHs can be analyzed with machine learning, and more specifically, a normalizing flow~\cite{2015arXiv150505770J, 2016arXiv160604934K, 2017arXiv170507057P} approach. We start by presenting our machine-learning model. Then we show the setup of our analysis and the results obtained. Finally, we give our conclusions and outline some prospects. 

\section{Machine learning for overlapping gravitational waves}
\label{sec:ML}

The use of machine learning in GW data analysis has been growing over the last years, having a wide range of applications. See Ref.~\cite{Cuoco:2020ogp} for a review. A subset of these methods fall under the umbrella of simulation-based inference~\cite{Cranmer:2020frontier}, and are being developed to perform parameter estimation for CBCs~\cite{Delaunoy:2020zcu, Green:2020hst, Green:2020dnx, Dax:2021myb, Dax:2021tsq,  Dax:2022pxd, Williams:2021qyt}. \cite{Dax:2021myb, Dax:2021tsq, Dax:2022pxd} use normalizing flows to get posterior distributions for BBH parameters. Such methods have been shown to have results close to those from MCMC and nested sampling. Our approach is somewhat similar to theirs, with some notable differences explained below. 

Our approach uses \emph{continuous conditional normalizing flows}~\cite{9089305, 2019arXiv191202762P}, which is a variant of normalizing flows suited for probabilistic modeling and Bayesian inference problems. Due to the recursive and continuous nature of these models their memory footprint can be quite small~\cite{chen2018neural}. These qualities allow for extensive training on home-grade GPUs while retaining the ability to capture complex distributions. We cover them in more detail below.

Normalizing flows are a method in machine learning through which a neural network can learn the mapping from some simple base distribution $p_u(\boldsymbol{u})$ (a Gaussian, for example) to a more complex final distribution $q(\boldsymbol{\theta})$. This is done through a series of invertible and differentiable transformations, summarized by a function $g(\boldsymbol{\theta})$. However, in our case, it is not sufficient to go from one distribution to the other. We also need to do this conditionally on the GW data that we wish to analyze. To account for this, we use \emph{conditional normalizing flows}~\cite{2019arXiv191200042W}, where the transformation functions are dependent on the data $\boldsymbol{d}$ (hence, $g = g(\boldsymbol{\theta}, \boldsymbol{d})$). A major difference with~\cite{2019arXiv191200042W} is that our base distributions are kept static; experiments on toymodels did not show any benefits in having conditional priors. Thus our model $g(\boldsymbol{\theta}, \boldsymbol{d})$ is a trainable conditional bijective function that transforms a simple 30-D Gaussian into a 30-D complex distribution. The bijectivity allows us to express and sample $q(\boldsymbol{\theta}|\boldsymbol{d})$ in terms of $g(\boldsymbol{\theta}, \boldsymbol{d})$ and $p_u(\boldsymbol{u})$ via the change of variable theorem:
\begin{equation}\label{eq:CdtNormalizingFlows}
    q(\boldsymbol{\theta} | \boldsymbol{d}) = \big| \rm{det}(J_{g^{-1}}(\boldsymbol{\theta}, \boldsymbol{d})) \big| p_u(g^{-1}(\boldsymbol{\theta}, \boldsymbol{d})) \, ,
\end{equation}
where $\rm{det}(J_{g^{-1}}(\boldsymbol{\theta}, \boldsymbol{d}))$ is the determinant of the Jacobian $J_{g^{-1}}(\boldsymbol{\theta}, \boldsymbol{d})$ of the transformation. We train the model by minimizing the forward KL-divergence, which is equivalent to maximum likelihood estimation~\cite{2017arXiv170507057P, 2016arXiv160506376P}. As noted by~\cite{Dax:2022pxd}, $q(\boldsymbol{\theta} | \boldsymbol{d})$ should cover the actual (Bayesian) posterior $p(\boldsymbol{\theta} | \boldsymbol{d})$, and asymptotically approach it as training progresses due to the mode-covering nature of the forward KL divergence. 

A distinctive choice of our method is the continuous nature of the flow, which is linked to the transformation function itself. Neural ordinary differential equations (neural ODEs)\cite{chen2018neural} are the foundation of continuous normalizing flows; neural ODEs are not represented by a stack of discrete layers but a hypernetwork \cite{ha2017hypernetworks}. Hypernetworks can be understood as regular networks where `external' inputs such as a (continuous) time or depth variable smoothly changes the output of the network for identical inputs. They can thus represent multiple networks or transformations. In \cite{chen2018neural}, hypernetworks are used to represent ODEs and are trained by using ODE-solvers and clever use of the adjoint sensitivity method. A continuous normalizing flow uses neural ODEs as it transformations. 

We will now explain the training of a continuous flow. For clarity we will use $h$ to refer to a continuous transformation and $g$ for a discrete one. If $\boldsymbol{\theta}(t)$ represents the samples from the distribution at a given time $t$, when going from $t_1$ to $t_2$, the continuous normalizing flow obeys an ordinary differential equation
\begin{equation}\label{eq:ContinuousFlow}
    \frac{\rm{d} \boldsymbol{\theta}(t)}{\rm{d}t} = h(t, \boldsymbol{\theta}(t)) \, .
\end{equation}
The change in likelihood associated with this `step' differs slightly from equations~\ref{eq:CdtNormalizingFlows} due to the continuous nature of the flow:
\begin{equation}
    \log (p(\boldsymbol{\theta}(t_1))) = \log (p(\boldsymbol{\theta}(t_0))) - \int_{t_0}^{t_1} \rm{Tr}\big[J_g(\boldsymbol{\theta}(t)) \big] \, .
\end{equation}
Assuming a non-stiff ODE the integration can be performed rapidly with state-of-the-art ode-solvers, which in our case is MALI \cite{zhuang2021mali}. In addition, we have to solve a trace instead of a determinant, which reduces the complexity, going from $\mathcal{O}(D^3)$ to at most $\mathcal{O}(D^2)$ with $D$ being the dimensionality of posterior space, and speed-ups the computation. Moreover, using continuous normalizing flows removes the need to use coupling layers between transformations, instead all parameter dimensions can be dependent on each other throughout the flow. Combining the continuous and conditional flows leads to continuous conditional normalizing flows, where the conditional consists of the GW $\boldsymbol{d}$ and the time $t$. 

We also need a better data representation than the raw strain to train and analyze the data. Therefore, we follow a similar approach as in~\cite{Green:2020dnx, Green:2020hst, Dax:2021myb, Dax:2021tsq, Dax:2022pxd}, using a single value decomposition (SVD)~\cite{2009arXiv0909.4061H} as summary statistics, reducing the dimension and the noise content of the data. Each of the 256 generated basis vectors is used as a kernel in 1D convolutions used as an initial layer in a residual convolutional neural network (CNN), enabling one to capture the time variance of the signal. Therefore, we do not need to use a Gibbs sampler to estimate the time of the signal as is done in~\cite{Dax:2021myb, Dax:2021tsq, Dax:2022pxd}, and can sample over time like any other variable which allows us to retain the likelihood of the samples. Furthermore, we also use a different representation for the angles. Instead of directly using their values, we project them onto a sphere for the sky location and onto a circle for the other angles. This makes for a better-posed domain for these angles, plays on the strong interpolation capacities of the network, and makes the training step easier.

In the end, our framework combines data representation as a hybrid between classical SVD and CNN, followed by the continuous normalizing flow network. It is worth noting that our entire framework is relatively small compared to the ones presented in~\cite{Dax:2021myb}. It enables the network to run on lower-end GPUs, but also means the network could be limited in his capacity to model the problem. 

\section{Data and setup}
\label{sec:framework}

To test the capacity of our framework to deal with overlapping signals, we start with a simplified setup. We consider a network made of the two LIGO detectors and the Virgo detector, at design sensitivity~\cite{aLIGOdesign, TheVirgo:2014hva}, and with a lower sensitive frequency of $20$Hz. We generate Gaussian noise from their power spectral density (PSD) and inject two precessing BBH mergers generated with the IMRPhenomPv2 waveform~\cite{Khan:2015jqa}. Our data frames have an 8 seconds duration. The chirp mass ($\mathcal{M}_c = (M_1 + M_2)^{3/5}/(M_1 M_2)^{1/5}$) is sampled from a uniform distribution between 10$M_{\odot}$ and 100$M_{\odot}$, and the mass ratio ($q = M_2 / M_1$) from a uniform distribution between 0.125 and 1. We constrain the individual component masses between 5$M_{\odot}$ and 100$M_{\odot}$. During the data generation, the luminosity distance is kept fixed. It is then rescaled to result in a network signal-to-noise ratio value taken randomly between 10 and 50, sampled from a beta distribution with central value 20. The time of coalescence for the two events is set randomly around a time of reference, with $t_{c} \in [t_{\rm{ref}} - 0.05, t_{\rm{ref}} + 0.05]$, ensuring that the two BBH merge in the high bias regime~\cite{Pizzati:2021apa}. The other parameters are drawn from their usual domain. Table~\ref{tab:parameters} gives an overview of the parameters and the function from which they are sampled. 

\begin{table}
    \centering
    \begin{tabular}{l r}
    \hline
    \hline
    Parameter & Function \\
    \hline
    \hline
         &  \\
    Chirp mass & $\mathcal{U}(10, 100)M_{\odot}$ \\
    Mass ratio & $\mathcal{U}(0.125, 1)$ \\
    Component masses & Constrained in [5, 100]$M_{\odot}$ \\
    Luminosity distance & Rescaled to follow SNR \\
    SNR & $\mathcal{B}(10, 50)$ \\
    Coalescence time & $\mathcal{U}(t_{\rm{ref}} - 0.05, t_{\rm{ref}} + 0.05)$ \\
    Spin Amplitudes & $\mathcal{U}(0, 1)$ \\
    Spin tilt angles & Uniform in sine \\
    Spin vector azimuthal angle &  $\mathcal{U}(0, 2\pi)$ \\
    Spin precession angle &  $\mathcal{U}(0, 2\pi)$ \\
    Inclination angle & Uniform in sine \\
    Wave polarization &  $\mathcal{U}(0, \pi)$ \\
    Phase of coalescence &  $\mathcal{U}(0, 2\pi)$ \\
    Right ascension &  $\mathcal{U}(0, 2\pi)$ \\
    Declination & Uniform in cosine \\
     & \\
     \hline
     \hline
    \end{tabular}
    \caption{Summary of the parameters considered for the BBHs generated as well as the functions used for their generation.}
    \label{tab:parameters}
\end{table}

During the training, we continuously generate the data by sampling the prior distributions for the events and generating a new noise realization for each data frame. The training is stopped when convergence is reached and before over-fitting occurs. Our model trained for about 12 days on a single Nvidia GeForce GTX 1080. 

\section{Results}
\label{sec:results}
\begin{figure*}
    \centering
    \includegraphics[keepaspectratio, width=0.49\textwidth]{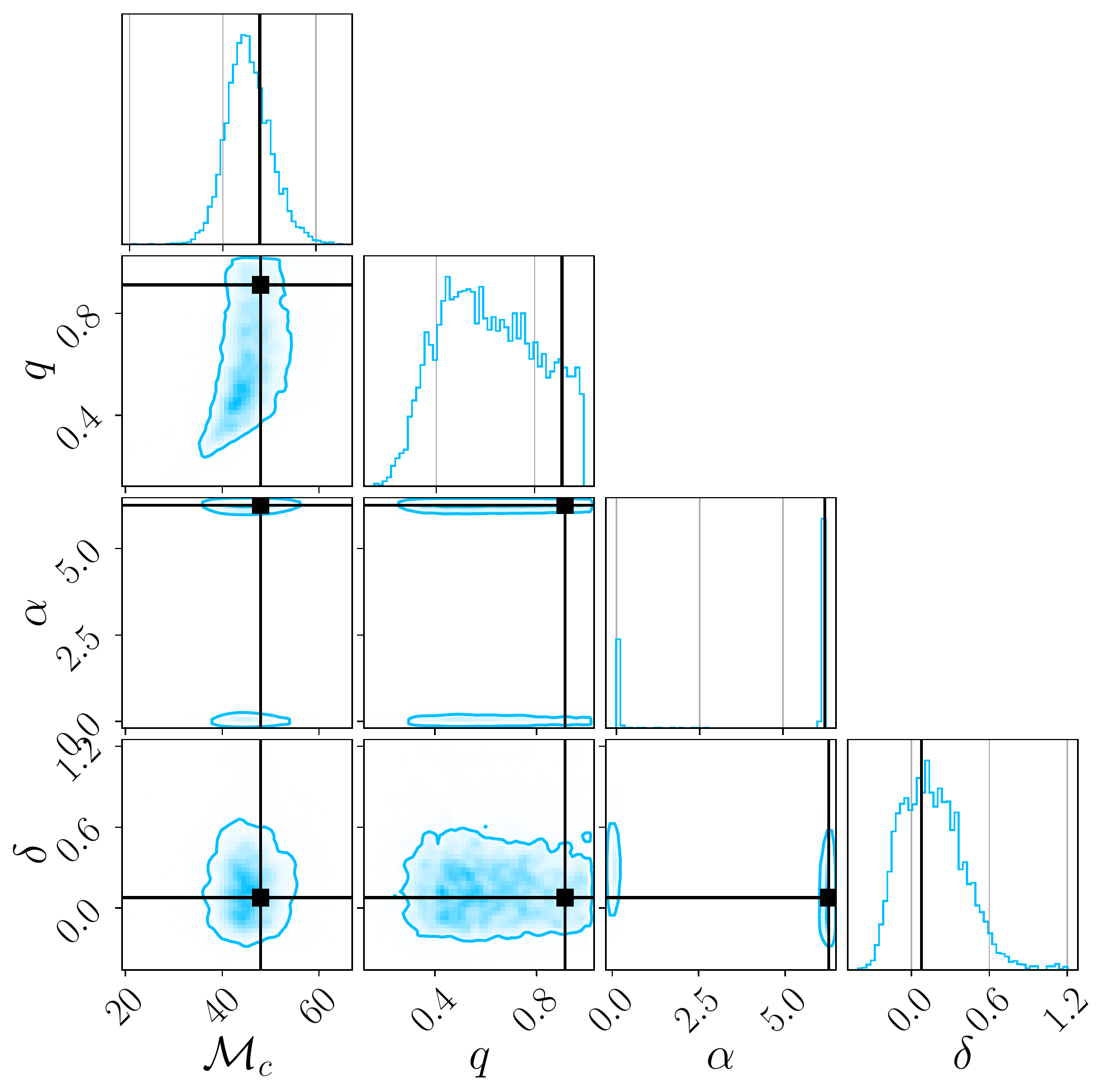}
    \includegraphics[keepaspectratio, width=0.49\textwidth]{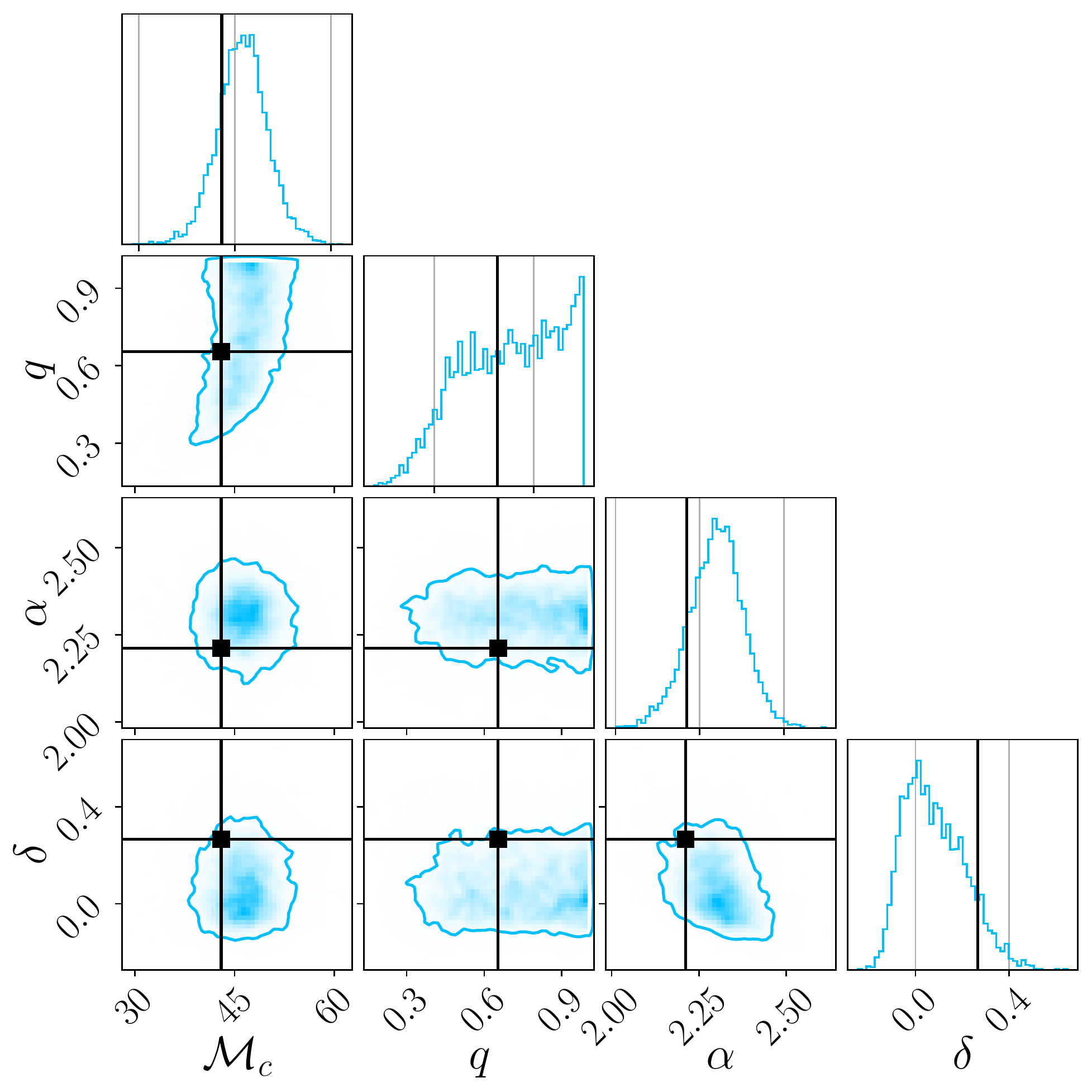}
    \caption{Example recovery for the masses and sky location of two overlapped BBHs with the machine learning approach. We see that the recovered values are well within the 90\% confidence interval.}
    \label{fig:CornerPlots}
\end{figure*}

First, we show the corner plots recovered for the masses and sky location of the two events in a pair in Fig.~\ref{fig:CornerPlots}. These are representative of our results. One can see that the injected values are within the 90\% confidence interval. This is the case for most events, regardless of the relative difference in arrival time or the SNR ratio between the two.

\begin{figure}
    \centering
    \includegraphics[keepaspectratio, width=0.5\textwidth]{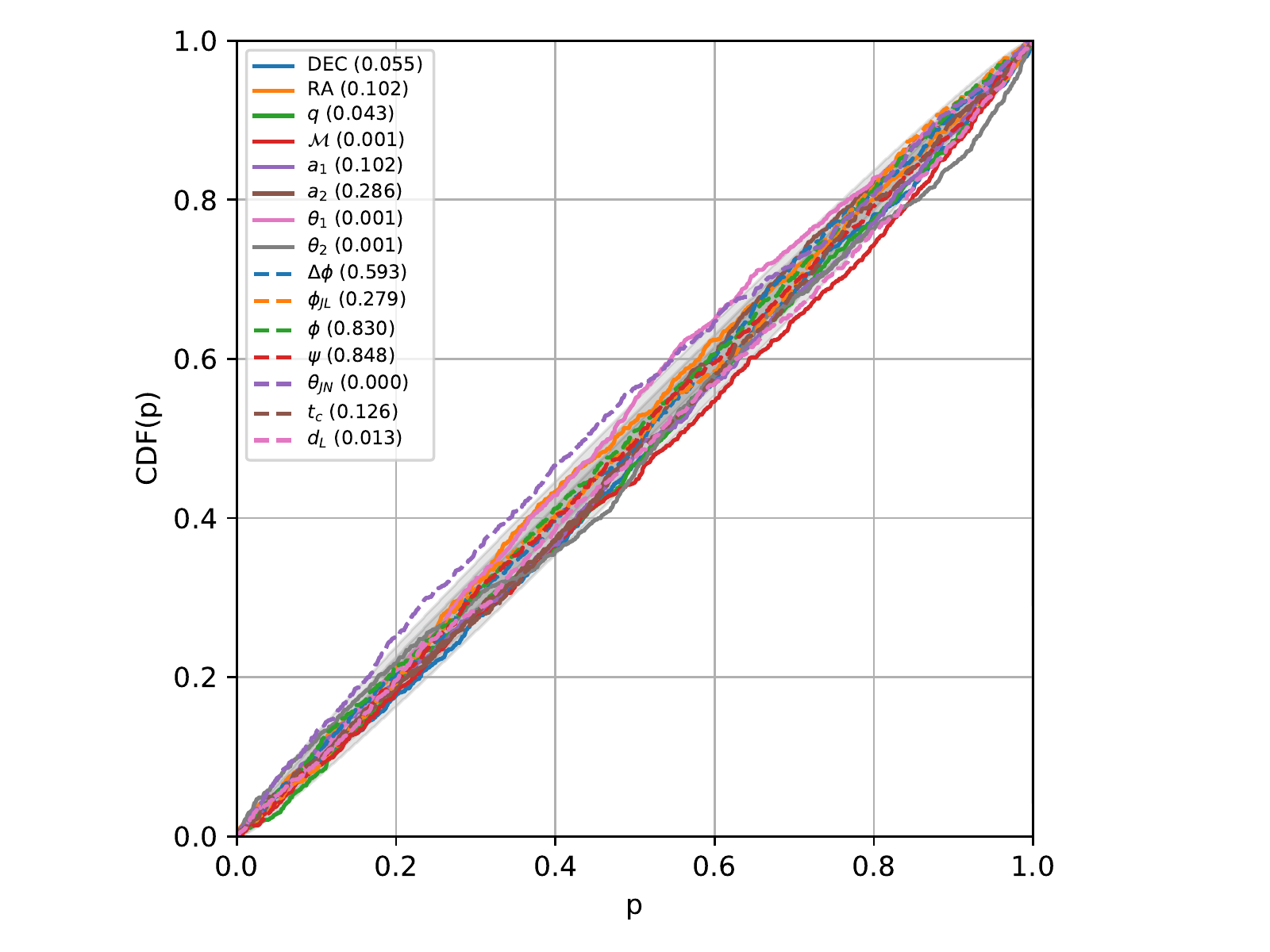}
    \includegraphics[keepaspectratio, width=0.5\textwidth]{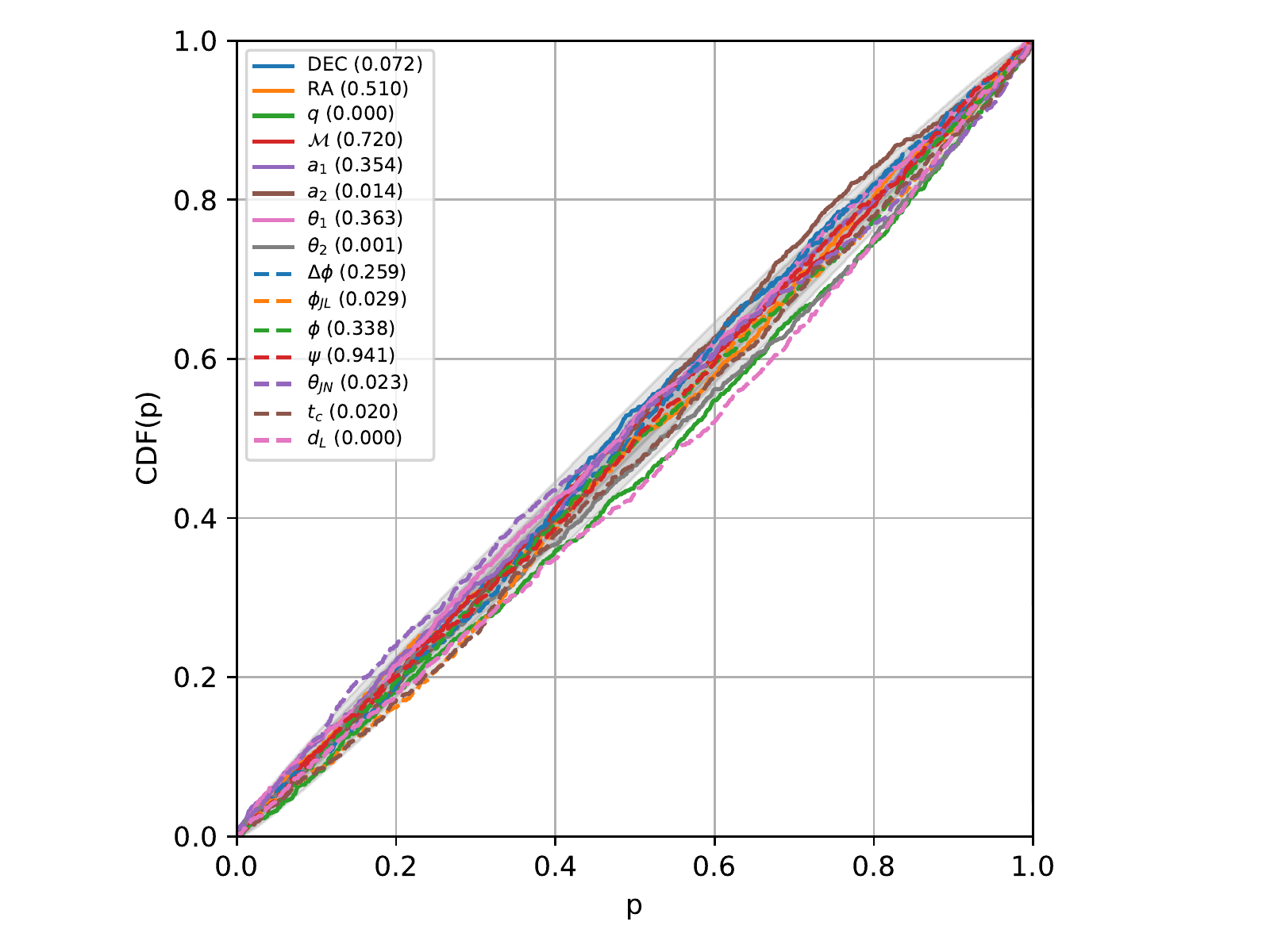}
    \caption{P-P plots for the recovered parameters for the two events in the data. In both cases, the lines align along the diagonal, showing that our method can be trusted. The larger spread can be due to bimodalities in the data, the increased complexity of the problem, and the reduced size of our network.}
    \label{fig:PPplot}
\end{figure}

To demonstrate the method's reliability, P-P plots for the two signals recovery are shown in  Fig.~\ref{fig:PPplot}. The P-P plot is constructed by sampling the posteriors of 1000 overlapped events. We then compute in which percentile of the distribution the injected value lies. If everything goes as expected the cumulative density functions should align along the diagonal. One can observe that this is the case for our network. Comparing this to the results given in~\cite{Dax:2021myb} for single signals, there is a broadening of the shell around the diagonal, showing more variability in the signal recovery. This means that our inference is less accurate than for single signals. Possible origins are the degenerate posteriors, increased complexity of the problem, and the reduced size of our network. In addition, an increased variability has been noted when going from single parameter estimation to joint parameter estimation methods in Bayesian approaches~\cite{Janquart:2022nyz}.

A trick used to alleviate the problem of degeneracies is to time-order the samples. Indeed, for two BBHs, the likelihood is symmetric in the two events. Therefore, the posteriors can get bimodal~\cite{Janquart:2022nyz}. While our training method formally labels one event as A and the other event as B, when the characteristics of the events are close, we may get somewhat bimodal posteriors. This probably also contributes to the higher variability of the P-P plot.

\begin{figure}
    \centering
    \includegraphics[keepaspectratio, width=0.49\textwidth]{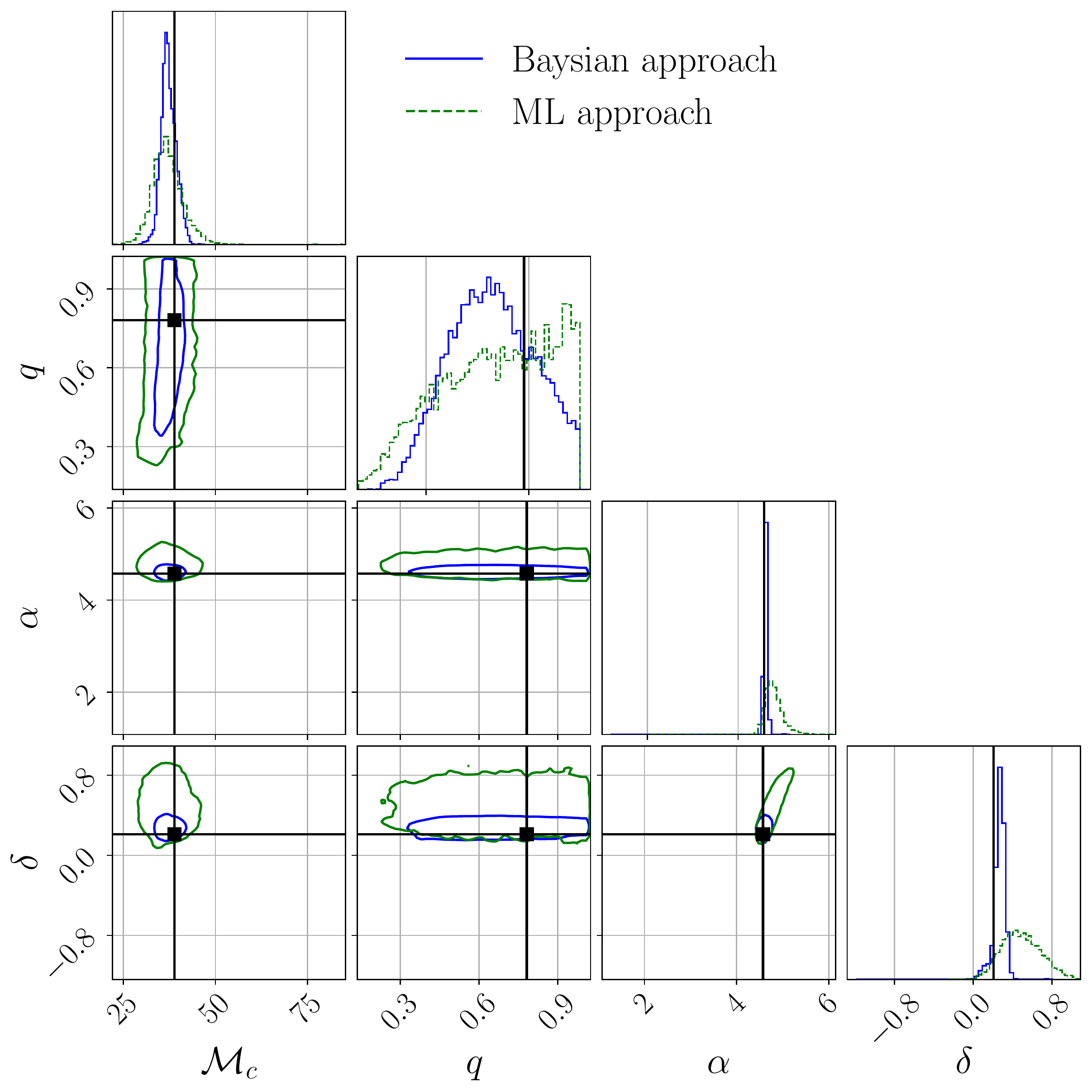}
    \includegraphics[keepaspectratio, width=0.49\textwidth]{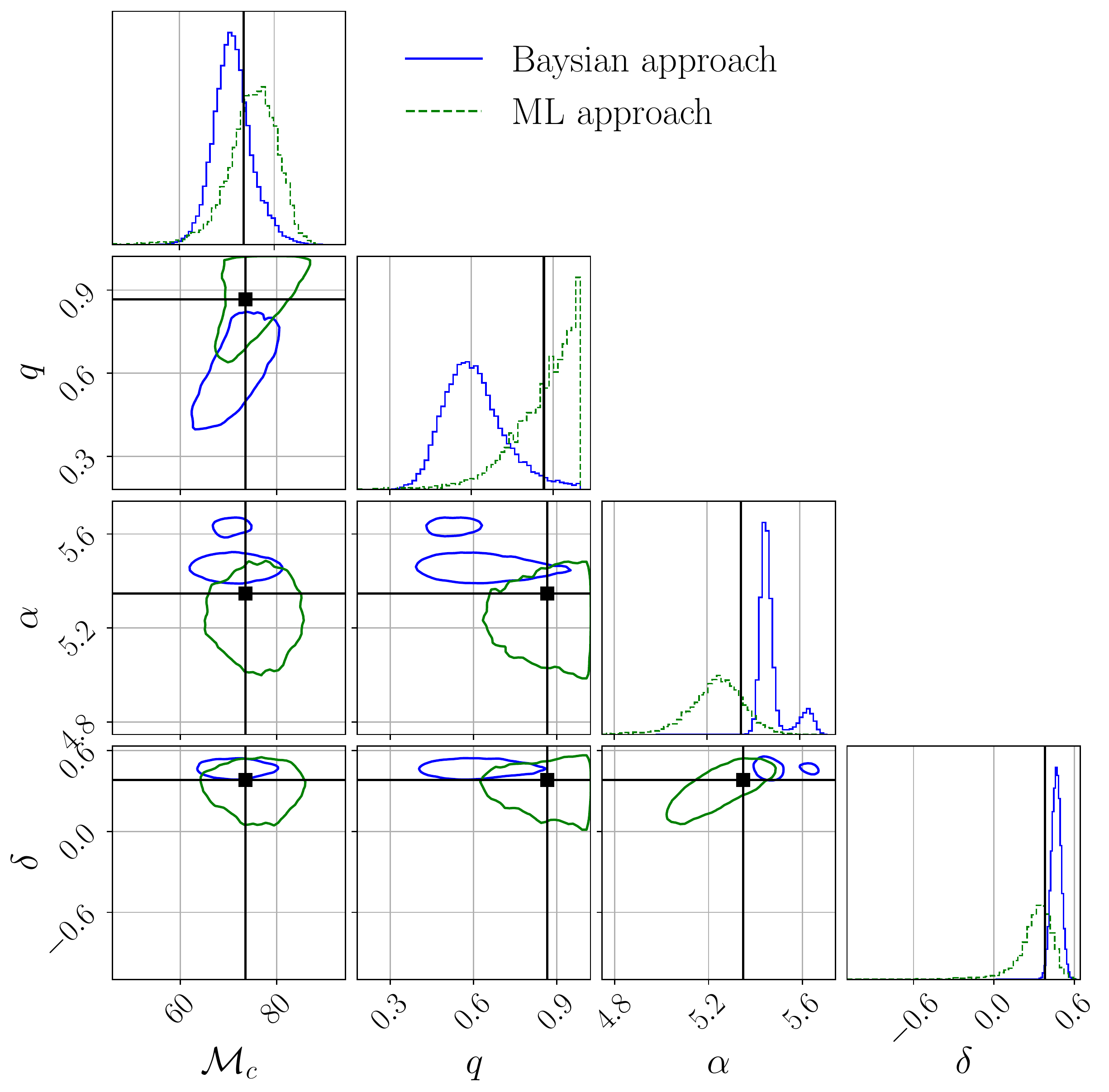}
    \caption{Comparison between our approach and the approach from~\cite{Janquart:2022nyz}. Our posteriors are generally broader but encapsulate the injected value within the 90\% confidence interval. This could be corrected by applying importance sampling on the samples given by our network.}
    \label{fig:compa_classical}
\end{figure}

Since parameter estimation for overlapping signals is still an active field of research, it is difficult to compare with traditional methods. While methods have been developed in~\cite{Janquart:2022nyz}, they are not yet fully stable and take a long time to analyze a BBH system. Therefore, making a statistically significant study comparing the two approaches seems a bit premature at this stage. However, to have some sense of the performances of our network compared to traditional methods, we make 15 injections complying with our network's setup and analyze them with the framework presented in~\cite{Janquart:2022nyz}. While this is not enough to make a statistical comparison, we can already identify some trends between the two pipelines. The first is that our machine pipeline typically has broader posteriors than the Bayesian approach. As mentioned in Ref.~\cite{Janquart:2022nyz}, the classical joint parameter estimation approach can sometimes get overconfident, where the recovered injected value lies outside of the 90\% confidence interval. Our method is not confronted with this bottleneck as the broader posterior encapsulates the injected value\footnote{In our 15 injections, we find 4 for which the Bayesian approach is overconfident. This is higher than values reported in~\cite{Janquart:2022nyz} and can be related to the closer merger times we are considering.}. Fig.~\ref{fig:compa_classical} represents the two situations: one where the Bayesian approach finds the event correctly, and one where we see that our ML approach covers the injected values while it does not for the classical approach. The origin of the larger posterior, which is not observed in the single parameter estimation machine learning-based methods, is probably due to the increased complexity of the problem combined with our network's small size. One possible avenue is applying importance sampling after the normalizing flow~\cite{Kolmus:2021buf, Dax:2022pxd}. This would increase the computational time, but the time needed to go from events to samples would still be well below the time taken by the traditional methods. However, such methods can be tricky, and additional modifications to our network could be needed.

Finally, an important advantage of our method is its speed. After being trained, it can analyze two overlapping BBH signals in about a second, to compare with $\mathcal{O}(20 days)$ reported in~\cite{Janquart:2022nyz}. While it is difficult to estimate the time gain for other types of signals, such as BNSs or NSBHs, we can expect the inference time after training not to be significantly larger than for BBHs. Since the computation time is a crucial aspect for studies in the 3G era, machine learning approaches seem to be more suited to study realistic scenarios for these detectors.

\section{Conclusions and Perspectives}
\label{sec:Conclusions}

In this work, we have presented a proof-of-concept machine learning-based method to analyze overlapping BBH signals. We focused on a 2G detector scenario with the two LIGO, and the Virgo detectors at design sensitivity, with a lower frequency cutoff of 20Hz. Our approach is based on continuous normalizing flows. 

While also using normalizing flows, as in~\cite{Green:2020dnx, Green:2020hst, Dax:2021myb, Dax:2021tsq, Dax:2022pxd}, we bring extra modifications that seem to help in the inference task. We represent the data through a mixture of SVD and convolutions, enabling us to sample directly over the events' arrival time, removing the need to use additional Gibbs sampling steps over that parameter and retaining the ability to access the likelihood of a sample. We also move to continuous conditional normalizing flows, reducing the computational cost of the method as we need to solve a trace instead of a determinant when going from one step to the other in the transformation. Finally, we also use a particular representation of the angles, projecting them onto circles (for the phase, the polarization, \dots) and spheres (for the sky location). We believe that these modifications make our network more supple, enabling it to deal with overlapping signals even in a reduced form.

With this simplified setup, we have shown that our approach is reliable, with posteriors consistent with the injected values. Our method takes about one week to train on a single GPU. After that, it only takes about a second to analyze two overlapped BBHs. While, in reality, other types of CBC mergers can happen, their inference after training should not be significantly longer than for BBHs. We also compared our machine learning method with classical Bayesian methods for overlapping signals. While our scheme leads to wider posteriors, it can correctly recover the injected values, even when the Baysian approach gets overconfident and misses the injection. A possibility to correct for the widened posteriors is to use importance sampling.

Our method's combined reliability and speed show that machine learning is a viable approach to analyze CBC mergers in the 3G era. More interestingly, it would even be possible without needing to account for the development of more powerful computational means and could enable some science-case studies for ET and CE soon. For example, once trained for all possible BBH systems, it could help study the BBH mass function in the 3G era. 

Still, one should note that extra improvements are needed before using our method in realistic 3G scenarios. One would first need to change our setup to the 3G detectors, a lower frequency and extreme SNRs that could be encountered. In addition, a wider range of objects should be accounted for. One should account for higher-order modes and eccentricity that could play a crucial role in the 3G era. Other modifications could also be implemented. Additionally, we need to account for the change in noise realization from one event to the other. Some of these steps, like changing the detectors, should be relatively easy. Others are more complex, as it is hard to perform parameter inference for long-lasting mergers due to the computational burden. So, extra developments in parameter estimation using machine learning would be required to get to the realistic 3G scenario. For overlapping signals, one would also benefit from developments in the classical study of the 3G scenario, such as how to deal with the noise characterization or the types of other events that could come into the data. 

In the end, there is still work to be done before machine learning can be used in realistic 3G scenarios. However, we believe that this work shows it is an interesting avenue and could be practical on a relatively short time scale.

\section*{Acknowledgements}
The authors are thankful to Tomasz Baka, Tom Heskes, and Twan van Laarhoven for discussions on related topics. The authors also thank Maximilian Dax for the careful re-reading of the manuscript. JL, JJ, and CVDB are supported by the research program of the Netherlands Organisation for Scientific Research (NWO). AK is supported by the NWO under the CORTEX project (NWA.1160.18.316). The authors are grateful for computational resources provided by the LIGO Laboratory and supported by the National Science Foundation Grants No. PHY-0757058 and No. PHY-0823459. We are grateful for computational resources provided by Cardiff University, and funded by an STFC grant supporting UK Involvement in the Operation of Advanced LIGO. 

\bibliography{bibliography}

\begin{thebibliography}{60}%
\makeatletter
\providecommand \@ifxundefined [1]{%
 \@ifx{#1\undefined}
}%
\providecommand \@ifnum [1]{%
 \ifnum #1\expandafter \@firstoftwo
 \else \expandafter \@secondoftwo
 \fi
}%
\providecommand \@ifx [1]{%
 \ifx #1\expandafter \@firstoftwo
 \else \expandafter \@secondoftwo
 \fi
}%
\providecommand \natexlab [1]{#1}%
\providecommand \enquote  [1]{``#1''}%
\providecommand \bibnamefont  [1]{#1}%
\providecommand \bibfnamefont [1]{#1}%
\providecommand \citenamefont [1]{#1}%
\providecommand \href@noop [0]{\@secondoftwo}%
\providecommand \href [0]{\begingroup \@sanitize@url \@href}%
\providecommand \@href[1]{\@@startlink{#1}\@@href}%
\providecommand \@@href[1]{\endgroup#1\@@endlink}%
\providecommand \@sanitize@url [0]{\catcode `\\12\catcode `\$12\catcode
  `\&12\catcode `\#12\catcode `\^12\catcode `\_12\catcode `\%12\relax}%
\providecommand \@@startlink[1]{}%
\providecommand \@@endlink[0]{}%
\providecommand \url  [0]{\begingroup\@sanitize@url \@url }%
\providecommand \@url [1]{\endgroup\@href {#1}{\urlprefix }}%
\providecommand \urlprefix  [0]{URL }%
\providecommand \Eprint [0]{\href }%
\providecommand \doibase [0]{http://dx.doi.org/}%
\providecommand \selectlanguage [0]{\@gobble}%
\providecommand \bibinfo  [0]{\@secondoftwo}%
\providecommand \bibfield  [0]{\@secondoftwo}%
\providecommand \translation [1]{[#1]}%
\providecommand \BibitemOpen [0]{}%
\providecommand \bibitemStop [0]{}%
\providecommand \bibitemNoStop [0]{.\EOS\space}%
\providecommand \EOS [0]{\spacefactor3000\relax}%
\providecommand \BibitemShut  [1]{\csname bibitem#1\endcsname}%
\let\auto@bib@innerbib\@empty
%</preamble>
\bibitem [{\citenamefont {Collaboration}(2015)}]{Aasi_2015}%
  \BibitemOpen
  \bibfield  {author} {\bibinfo {author} {\bibfnamefont {T.~L.~S.}\
  \bibnamefont {Collaboration}},\ }\href {\doibase
  10.1088/0264-9381/32/7/074001} {\bibfield  {journal} {\bibinfo  {journal}
  {Classical and Quantum Gravity}\ }\textbf {\bibinfo {volume} {32}},\ \bibinfo
  {pages} {074001} (\bibinfo {year} {2015})}\BibitemShut {NoStop}%
\bibitem [{\citenamefont {Acernese}\ \emph {et~al.}(2015)\citenamefont
  {Acernese} \emph {et~al.}}]{TheVirgo:2014hva}%
  \BibitemOpen
  \bibfield  {author} {\bibinfo {author} {\bibfnamefont {F.}~\bibnamefont
  {Acernese}} \emph {et~al.} (\bibinfo {collaboration} {VIRGO}),\ }\href
  {\doibase 10.1088/0264-9381/32/2/024001} {\bibfield  {journal} {\bibinfo
  {journal} {Class. Quant. Grav.}\ }\textbf {\bibinfo {volume} {32}},\ \bibinfo
  {pages} {024001} (\bibinfo {year} {2015})},\ \Eprint
  {http://arxiv.org/abs/1408.3978} {arXiv:1408.3978 [gr-qc]} \BibitemShut
  {NoStop}%
\bibitem [{\citenamefont {Abbott}\ \emph
  {et~al.}(2021{\natexlab{a}})\citenamefont {Abbott} \emph
  {et~al.}}]{LIGOScientific:2021djp}%
  \BibitemOpen
  \bibfield  {author} {\bibinfo {author} {\bibfnamefont {R.}~\bibnamefont
  {Abbott}} \emph {et~al.} (\bibinfo {collaboration} {LIGO Scientific, VIRGO,
  KAGRA}),\ }\href@noop {} {\  (\bibinfo {year} {2021}{\natexlab{a}})},\
  \Eprint {http://arxiv.org/abs/2111.03606} {arXiv:2111.03606 [gr-qc]}
  \BibitemShut {NoStop}%
\bibitem [{\citenamefont {Abbott}\ \emph
  {et~al.}(2021{\natexlab{b}})\citenamefont {Abbott} \emph
  {et~al.}}]{LIGOScientific:2021sio}%
  \BibitemOpen
  \bibfield  {author} {\bibinfo {author} {\bibfnamefont {R.}~\bibnamefont
  {Abbott}} \emph {et~al.} (\bibinfo {collaboration} {LIGO Scientific, VIRGO,
  KAGRA}),\ }\href@noop {} {\  (\bibinfo {year} {2021}{\natexlab{b}})},\
  \Eprint {http://arxiv.org/abs/2112.06861} {arXiv:2112.06861 [gr-qc]}
  \BibitemShut {NoStop}%
\bibitem [{\citenamefont {Abbott}\ \emph
  {et~al.}(2021{\natexlab{c}})\citenamefont {Abbott} \emph
  {et~al.}}]{LIGOScientific:2021psn}%
  \BibitemOpen
  \bibfield  {author} {\bibinfo {author} {\bibfnamefont {R.}~\bibnamefont
  {Abbott}} \emph {et~al.} (\bibinfo {collaboration} {LIGO Scientific, VIRGO,
  KAGRA}),\ }\href@noop {} {\  (\bibinfo {year} {2021}{\natexlab{c}})},\
  \Eprint {http://arxiv.org/abs/2111.03634} {arXiv:2111.03634 [astro-ph.HE]}
  \BibitemShut {NoStop}%
\bibitem [{\citenamefont {Abbott}\ \emph
  {et~al.}(2021{\natexlab{d}})\citenamefont {Abbott} \emph
  {et~al.}}]{LIGOScientific:2021aug}%
  \BibitemOpen
  \bibfield  {author} {\bibinfo {author} {\bibfnamefont {R.}~\bibnamefont
  {Abbott}} \emph {et~al.} (\bibinfo {collaboration} {LIGO Scientific, VIRGO,
  KAGRA}),\ }\href@noop {} {\  (\bibinfo {year} {2021}{\natexlab{d}})},\
  \Eprint {http://arxiv.org/abs/2111.03604} {arXiv:2111.03604 [astro-ph.CO]}
  \BibitemShut {NoStop}%
\bibitem [{\citenamefont {Somiya}(2012)}]{Somiya:2011np}%
  \BibitemOpen
  \bibfield  {author} {\bibinfo {author} {\bibfnamefont {K.}~\bibnamefont
  {Somiya}} (\bibinfo {collaboration} {KAGRA}),\ }\href {\doibase
  10.1088/0264-9381/29/12/124007} {\bibfield  {journal} {\bibinfo  {journal}
  {Class. Quant. Grav.}\ }\textbf {\bibinfo {volume} {29}},\ \bibinfo {pages}
  {124007} (\bibinfo {year} {2012})},\ \Eprint {http://arxiv.org/abs/1111.7185}
  {arXiv:1111.7185 [gr-qc]} \BibitemShut {NoStop}%
\bibitem [{\citenamefont {Aso}\ \emph {et~al.}(2013)\citenamefont {Aso},
  \citenamefont {Michimura}, \citenamefont {Somiya}, \citenamefont {Ando},
  \citenamefont {Miyakawa}, \citenamefont {Sekiguchi}, \citenamefont
  {Tatsumi},\ and\ \citenamefont {Yamamoto}}]{Aso:2013eba}%
  \BibitemOpen
  \bibfield  {author} {\bibinfo {author} {\bibfnamefont {Y.}~\bibnamefont
  {Aso}}, \bibinfo {author} {\bibfnamefont {Y.}~\bibnamefont {Michimura}},
  \bibinfo {author} {\bibfnamefont {K.}~\bibnamefont {Somiya}}, \bibinfo
  {author} {\bibfnamefont {M.}~\bibnamefont {Ando}}, \bibinfo {author}
  {\bibfnamefont {O.}~\bibnamefont {Miyakawa}}, \bibinfo {author}
  {\bibfnamefont {T.}~\bibnamefont {Sekiguchi}}, \bibinfo {author}
  {\bibfnamefont {D.}~\bibnamefont {Tatsumi}}, \ and\ \bibinfo {author}
  {\bibfnamefont {H.}~\bibnamefont {Yamamoto}} (\bibinfo {collaboration}
  {KAGRA}),\ }\href {\doibase 10.1103/PhysRevD.88.043007} {\bibfield  {journal}
  {\bibinfo  {journal} {Phys. Rev. D}\ }\textbf {\bibinfo {volume} {88}},\
  \bibinfo {pages} {043007} (\bibinfo {year} {2013})},\ \Eprint
  {http://arxiv.org/abs/1306.6747} {arXiv:1306.6747 [gr-qc]} \BibitemShut
  {NoStop}%
\bibitem [{\citenamefont {Akutsu}\ \emph {et~al.}(2019)\citenamefont {Akutsu}
  \emph {et~al.}}]{Akutsu:2018axf}%
  \BibitemOpen
  \bibfield  {author} {\bibinfo {author} {\bibfnamefont {T.}~\bibnamefont
  {Akutsu}} \emph {et~al.} (\bibinfo {collaboration} {KAGRA}),\ }\href
  {\doibase 10.1038/s41550-018-0658-y} {\bibfield  {journal} {\bibinfo
  {journal} {Nature Astron.}\ }\textbf {\bibinfo {volume} {3}},\ \bibinfo
  {pages} {35} (\bibinfo {year} {2019})},\ \Eprint
  {http://arxiv.org/abs/1811.08079} {arXiv:1811.08079 [gr-qc]} \BibitemShut
  {NoStop}%
\bibitem [{\citenamefont {Akutsu}\ \emph {et~al.}(2020)\citenamefont {Akutsu},
  \citenamefont {Ando}, \citenamefont {Arai} \emph {et~al.}}]{Akutsu:2020his}%
  \BibitemOpen
  \bibfield  {author} {\bibinfo {author} {\bibfnamefont {T.}~\bibnamefont
  {Akutsu}}, \bibinfo {author} {\bibfnamefont {M.}~\bibnamefont {Ando}},
  \bibinfo {author} {\bibfnamefont {K.}~\bibnamefont {Arai}},  \emph {et~al.},\
  }\href@noop {} {\bibfield  {journal} {\bibinfo  {journal} {arXiv e-prints}\
  ,\ \bibinfo {eid} {arXiv:2005.05574}} (\bibinfo {year} {2020})},\ \Eprint
  {http://arxiv.org/abs/2005.05574} {arXiv:2005.05574 [physics.ins-det]}
  \BibitemShut {NoStop}%
\bibitem [{\citenamefont {Akutsu}\ \emph {et~al.}(2021)\citenamefont {Akutsu}
  \emph {et~al.}}]{KAGRA:2020tym}%
  \BibitemOpen
  \bibfield  {author} {\bibinfo {author} {\bibfnamefont {T.}~\bibnamefont
  {Akutsu}} \emph {et~al.} (\bibinfo {collaboration} {KAGRA}),\ }\href
  {\doibase 10.1093/ptep/ptaa125} {\bibfield  {journal} {\bibinfo  {journal}
  {PTEP}\ }\textbf {\bibinfo {volume} {2021}},\ \bibinfo {pages} {05A101}
  (\bibinfo {year} {2021})},\ \Eprint {http://arxiv.org/abs/2005.05574}
  {arXiv:2005.05574 [physics.ins-det]} \BibitemShut {NoStop}%
\bibitem [{\citenamefont {Iyer}\ \emph {et~al.}(2011)\citenamefont {Iyer},
  \citenamefont {Souradeep}, \citenamefont {Unnikrishnan}, \citenamefont
  {Dhurandhar}, \citenamefont {Raja},\ and\ \citenamefont
  {Sengupta}}]{LigoIndia}%
  \BibitemOpen
  \bibfield  {author} {\bibinfo {author} {\bibfnamefont {B.}~\bibnamefont
  {Iyer}}, \bibinfo {author} {\bibfnamefont {T.}~\bibnamefont {Souradeep}},
  \bibinfo {author} {\bibfnamefont {C.}~\bibnamefont {Unnikrishnan}}, \bibinfo
  {author} {\bibfnamefont {S.}~\bibnamefont {Dhurandhar}}, \bibinfo {author}
  {\bibfnamefont {S.}~\bibnamefont {Raja}}, \ and\ \bibinfo {author}
  {\bibfnamefont {A.}~\bibnamefont {Sengupta}},\ }\href@noop {} {\enquote
  {\bibinfo {title} {{LIGO-India Tech. rep. }},}\ }\bibinfo {howpublished}
  {\url{https://dcc.ligo.org/LIGO-M1100296/public}} (\bibinfo {year}
  {2011})\BibitemShut {NoStop}%
\bibitem [{\citenamefont {Punturo}\ \emph {et~al.}(2010)\citenamefont {Punturo}
  \emph {et~al.}}]{Punturo_2010}%
  \BibitemOpen
  \bibfield  {author} {\bibinfo {author} {\bibfnamefont {M.}~\bibnamefont
  {Punturo}} \emph {et~al.},\ }\href {\doibase 10.1088/0264-9381/27/19/194002}
  {\bibfield  {journal} {\bibinfo  {journal} {Classical and Quantum Gravity}\
  }\textbf {\bibinfo {volume} {27}},\ \bibinfo {pages} {194002} (\bibinfo
  {year} {2010})}\BibitemShut {NoStop}%
\bibitem [{\citenamefont {Hild}\ \emph {et~al.}(2011)\citenamefont {Hild} \emph
  {et~al.}}]{Hild:2010id}%
  \BibitemOpen
  \bibfield  {author} {\bibinfo {author} {\bibfnamefont {S.}~\bibnamefont
  {Hild}} \emph {et~al.},\ }\href {\doibase 10.1088/0264-9381/28/9/094013}
  {\bibfield  {journal} {\bibinfo  {journal} {Class. Quant. Grav.}\ }\textbf
  {\bibinfo {volume} {28}},\ \bibinfo {pages} {094013} (\bibinfo {year}
  {2011})},\ \Eprint {http://arxiv.org/abs/1012.0908} {arXiv:1012.0908 [gr-qc]}
  \BibitemShut {NoStop}%
%%CITATION = ARXIV:1012.0908;%%
\bibitem [{\citenamefont {Reitze}\ \emph {et~al.}(2019)\citenamefont {Reitze}
  \emph {et~al.}}]{Reitze:2019iox}%
  \BibitemOpen
  \bibfield  {author} {\bibinfo {author} {\bibfnamefont {D.}~\bibnamefont
  {Reitze}} \emph {et~al.},\ }\href@noop {} {\bibfield  {journal} {\bibinfo
  {journal} {Bull. Am. Astron. Soc.}\ }\textbf {\bibinfo {volume} {51}},\
  \bibinfo {pages} {035} (\bibinfo {year} {2019})},\ \Eprint
  {http://arxiv.org/abs/1907.04833} {arXiv:1907.04833 [astro-ph.IM]}
  \BibitemShut {NoStop}%
\bibitem [{\citenamefont {Abbott}\ \emph {et~al.}(2017)\citenamefont {Abbott},
  \citenamefont {Abbott}, \citenamefont {Abbott}, \citenamefont {Abernathy},
  \citenamefont {Ackley},\ and\ \citenamefont {Adams}}]{Abbott_2017}%
  \BibitemOpen
  \bibfield  {author} {\bibinfo {author} {\bibfnamefont {B.~P.}\ \bibnamefont
  {Abbott}}, \bibinfo {author} {\bibfnamefont {R.}~\bibnamefont {Abbott}},
  \bibinfo {author} {\bibfnamefont {T.~D.}\ \bibnamefont {Abbott}}, \bibinfo
  {author} {\bibfnamefont {M.~R.}\ \bibnamefont {Abernathy}}, \bibinfo {author}
  {\bibfnamefont {K.}~\bibnamefont {Ackley}}, \ and\ \bibinfo {author}
  {\bibfnamefont {C.}~\bibnamefont {Adams}},\ }\href {\doibase
  10.1088/1361-6382/aa51f4} {\bibfield  {journal} {\bibinfo  {journal}
  {Classical and Quantum Gravity}\ }\textbf {\bibinfo {volume} {34}},\ \bibinfo
  {pages} {044001} (\bibinfo {year} {2017})}\BibitemShut {NoStop}%
\bibitem [{\citenamefont {Regimbau}\ \emph {et~al.}(2017)\citenamefont
  {Regimbau}, \citenamefont {Evans}, \citenamefont {Christensen}, \citenamefont
  {Katsavounidis}, \citenamefont {Sathyaprakash},\ and\ \citenamefont
  {Vitale}}]{PhysRevLett.118.151105}%
  \BibitemOpen
  \bibfield  {author} {\bibinfo {author} {\bibfnamefont {T.}~\bibnamefont
  {Regimbau}}, \bibinfo {author} {\bibfnamefont {M.}~\bibnamefont {Evans}},
  \bibinfo {author} {\bibfnamefont {N.}~\bibnamefont {Christensen}}, \bibinfo
  {author} {\bibfnamefont {E.}~\bibnamefont {Katsavounidis}}, \bibinfo {author}
  {\bibfnamefont {B.}~\bibnamefont {Sathyaprakash}}, \ and\ \bibinfo {author}
  {\bibfnamefont {S.}~\bibnamefont {Vitale}},\ }\href {\doibase
  10.1103/PhysRevLett.118.151105} {\bibfield  {journal} {\bibinfo  {journal}
  {Phys. Rev. Lett.}\ }\textbf {\bibinfo {volume} {118}},\ \bibinfo {pages}
  {151105} (\bibinfo {year} {2017})}\BibitemShut {NoStop}%
\bibitem [{\citenamefont {Sathyaprakash}\ \emph {et~al.}(2012)\citenamefont
  {Sathyaprakash} \emph {et~al.}}]{Sathyaprakash:2012jk}%
  \BibitemOpen
  \bibfield  {author} {\bibinfo {author} {\bibfnamefont {B.}~\bibnamefont
  {Sathyaprakash}} \emph {et~al.},\ }\href {\doibase
  10.1088/0264-9381/29/12/124013} {\bibfield  {journal} {\bibinfo  {journal}
  {Class. Quant. Grav.}\ }\textbf {\bibinfo {volume} {29}},\ \bibinfo {pages}
  {124013} (\bibinfo {year} {2012})},\ \bibinfo {note} {[Erratum:
  Class.Quant.Grav. 30, 079501 (2013)]},\ \Eprint
  {http://arxiv.org/abs/1206.0331} {arXiv:1206.0331 [gr-qc]} \BibitemShut
  {NoStop}%
\bibitem [{\citenamefont {Regimbau}\ and\ \citenamefont
  {Hughes}(2009)}]{Regimbau:2009rk}%
  \BibitemOpen
  \bibfield  {author} {\bibinfo {author} {\bibfnamefont {T.}~\bibnamefont
  {Regimbau}}\ and\ \bibinfo {author} {\bibfnamefont {S.~A.}\ \bibnamefont
  {Hughes}},\ }\href {\doibase 10.1103/PhysRevD.79.062002} {\bibfield
  {journal} {\bibinfo  {journal} {Phys. Rev. D}\ }\textbf {\bibinfo {volume}
  {79}},\ \bibinfo {pages} {062002} (\bibinfo {year} {2009})},\ \Eprint
  {http://arxiv.org/abs/0901.2958} {arXiv:0901.2958 [gr-qc]} \BibitemShut
  {NoStop}%
\bibitem [{\citenamefont {Samajdar}\ \emph {et~al.}(2021)\citenamefont
  {Samajdar}, \citenamefont {Janquart}, \citenamefont {Van Den~Broeck},\ and\
  \citenamefont {Dietrich}}]{Samajdar:2021egv}%
  \BibitemOpen
  \bibfield  {author} {\bibinfo {author} {\bibfnamefont {A.}~\bibnamefont
  {Samajdar}}, \bibinfo {author} {\bibfnamefont {J.}~\bibnamefont {Janquart}},
  \bibinfo {author} {\bibfnamefont {C.}~\bibnamefont {Van Den~Broeck}}, \ and\
  \bibinfo {author} {\bibfnamefont {T.}~\bibnamefont {Dietrich}},\ }\href
  {\doibase 10.1103/PhysRevD.104.044003} {\bibfield  {journal} {\bibinfo
  {journal} {Phys. Rev. D}\ }\textbf {\bibinfo {volume} {104}},\ \bibinfo
  {pages} {044003} (\bibinfo {year} {2021})},\ \Eprint
  {http://arxiv.org/abs/2102.07544} {arXiv:2102.07544 [gr-qc]} \BibitemShut
  {NoStop}%
\bibitem [{\citenamefont {Pizzati}\ \emph {et~al.}(2022)\citenamefont
  {Pizzati}, \citenamefont {Sachdev}, \citenamefont {Gupta},\ and\
  \citenamefont {Sathyaprakash}}]{Pizzati:2021apa}%
  \BibitemOpen
  \bibfield  {author} {\bibinfo {author} {\bibfnamefont {E.}~\bibnamefont
  {Pizzati}}, \bibinfo {author} {\bibfnamefont {S.}~\bibnamefont {Sachdev}},
  \bibinfo {author} {\bibfnamefont {A.}~\bibnamefont {Gupta}}, \ and\ \bibinfo
  {author} {\bibfnamefont {B.}~\bibnamefont {Sathyaprakash}},\ }\href {\doibase
  10.1103/PhysRevD.105.104016} {\bibfield  {journal} {\bibinfo  {journal}
  {Phys. Rev. D}\ }\textbf {\bibinfo {volume} {105}},\ \bibinfo {pages}
  {104016} (\bibinfo {year} {2022})},\ \Eprint
  {http://arxiv.org/abs/2102.07692} {arXiv:2102.07692 [gr-qc]} \BibitemShut
  {NoStop}%
\bibitem [{\citenamefont {Relton}\ and\ \citenamefont
  {Raymond}(2021)}]{Relton:2021cax}%
  \BibitemOpen
  \bibfield  {author} {\bibinfo {author} {\bibfnamefont {P.}~\bibnamefont
  {Relton}}\ and\ \bibinfo {author} {\bibfnamefont {V.}~\bibnamefont
  {Raymond}},\ }\href {\doibase 10.1103/PhysRevD.104.084039} {\bibfield
  {journal} {\bibinfo  {journal} {Phys. Rev. D}\ }\textbf {\bibinfo {volume}
  {104}},\ \bibinfo {pages} {084039} (\bibinfo {year} {2021})},\ \Eprint
  {http://arxiv.org/abs/2103.16225} {arXiv:2103.16225 [gr-qc]} \BibitemShut
  {NoStop}%
\bibitem [{\citenamefont {Himemoto}\ \emph {et~al.}(2021)\citenamefont
  {Himemoto}, \citenamefont {Nishizawa},\ and\ \citenamefont
  {Taruya}}]{Himemoto:2021ukb}%
  \BibitemOpen
  \bibfield  {author} {\bibinfo {author} {\bibfnamefont {Y.}~\bibnamefont
  {Himemoto}}, \bibinfo {author} {\bibfnamefont {A.}~\bibnamefont {Nishizawa}},
  \ and\ \bibinfo {author} {\bibfnamefont {A.}~\bibnamefont {Taruya}},\ }\href
  {\doibase 10.1103/PhysRevD.104.044010} {\bibfield  {journal} {\bibinfo
  {journal} {Phys. Rev. D}\ }\textbf {\bibinfo {volume} {104}},\ \bibinfo
  {pages} {044010} (\bibinfo {year} {2021})},\ \Eprint
  {http://arxiv.org/abs/2103.14816} {arXiv:2103.14816 [gr-qc]} \BibitemShut
  {NoStop}%
\bibitem [{\citenamefont {Antonelli}\ \emph {et~al.}(2021)\citenamefont
  {Antonelli}, \citenamefont {Burke},\ and\ \citenamefont
  {Gair}}]{Antonelli:2021vwg}%
  \BibitemOpen
  \bibfield  {author} {\bibinfo {author} {\bibfnamefont {A.}~\bibnamefont
  {Antonelli}}, \bibinfo {author} {\bibfnamefont {O.}~\bibnamefont {Burke}}, \
  and\ \bibinfo {author} {\bibfnamefont {J.~R.}\ \bibnamefont {Gair}},\ }\href
  {\doibase 10.1093/mnras/stab2358} {\bibfield  {journal} {\bibinfo  {journal}
  {Mon. Not. Roy. Astron. Soc.}\ }\textbf {\bibinfo {volume} {507}},\ \bibinfo
  {pages} {5069} (\bibinfo {year} {2021})},\ \Eprint
  {http://arxiv.org/abs/2104.01897} {arXiv:2104.01897 [gr-qc]} \BibitemShut
  {NoStop}%
\bibitem [{\citenamefont {Wu}\ and\ \citenamefont {Nitz}(2022)}]{Wu:2022pyg}%
  \BibitemOpen
  \bibfield  {author} {\bibinfo {author} {\bibfnamefont {S.}~\bibnamefont
  {Wu}}\ and\ \bibinfo {author} {\bibfnamefont {A.~H.}\ \bibnamefont {Nitz}},\
  }\href@noop {} {\  (\bibinfo {year} {2022})},\ \Eprint
  {http://arxiv.org/abs/2209.03135} {arXiv:2209.03135 [astro-ph.IM]}
  \BibitemShut {NoStop}%
\bibitem [{\citenamefont {Sachdev}\ \emph {et~al.}(2020)\citenamefont
  {Sachdev}, \citenamefont {Regimbau},\ and\ \citenamefont
  {Sathyaprakash}}]{Sachdev2020:pol}%
  \BibitemOpen
  \bibfield  {author} {\bibinfo {author} {\bibfnamefont {S.}~\bibnamefont
  {Sachdev}}, \bibinfo {author} {\bibfnamefont {T.}~\bibnamefont {Regimbau}}, \
  and\ \bibinfo {author} {\bibfnamefont {B.~S.}\ \bibnamefont
  {Sathyaprakash}},\ }\href {\doibase 10.1103/PhysRevD.102.024051} {\bibfield
  {journal} {\bibinfo  {journal} {Phys. Rev. D}\ }\textbf {\bibinfo {volume}
  {102}},\ \bibinfo {pages} {024051} (\bibinfo {year} {2020})}\BibitemShut
  {NoStop}%
\bibitem [{\citenamefont {Sharma}\ and\ \citenamefont
  {Harms}(2020)}]{Sharma:2022kds}%
  \BibitemOpen
  \bibfield  {author} {\bibinfo {author} {\bibfnamefont {A.}~\bibnamefont
  {Sharma}}\ and\ \bibinfo {author} {\bibfnamefont {J.}~\bibnamefont {Harms}},\
  }\href {\doibase 10.1103/PhysRevD.102.063009} {\bibfield  {journal} {\bibinfo
   {journal} {Phys. Rev. D}\ }\textbf {\bibinfo {volume} {102}},\ \bibinfo
  {pages} {063009} (\bibinfo {year} {2020})}\BibitemShut {NoStop}%
\bibitem [{\citenamefont {Biscoveanu}\ \emph {et~al.}(2020)\citenamefont
  {Biscoveanu}, \citenamefont {Talbot}, \citenamefont {Thrane},\ and\
  \citenamefont {Smith}}]{Biscoveanu:2020ste}%
  \BibitemOpen
  \bibfield  {author} {\bibinfo {author} {\bibfnamefont {S.}~\bibnamefont
  {Biscoveanu}}, \bibinfo {author} {\bibfnamefont {C.}~\bibnamefont {Talbot}},
  \bibinfo {author} {\bibfnamefont {E.}~\bibnamefont {Thrane}}, \ and\ \bibinfo
  {author} {\bibfnamefont {R.}~\bibnamefont {Smith}},\ }\href {\doibase
  10.1103/PhysRevLett.125.241101} {\bibfield  {journal} {\bibinfo  {journal}
  {Phys. Rev. Lett.}\ }\textbf {\bibinfo {volume} {125}},\ \bibinfo {pages}
  {241101} (\bibinfo {year} {2020})}\BibitemShut {NoStop}%
\bibitem [{\citenamefont {Zhou}\ \emph
  {et~al.}(2022{\natexlab{a}})\citenamefont {Zhou}, \citenamefont {Reali},
  \citenamefont {Berti}, \citenamefont {\c{C}al\i{}\c{s}kan}, \citenamefont
  {Creque-Sarbinowski}, \citenamefont {Kamionkowski},\ and\ \citenamefont
  {Sathyaprakash}}]{Zhou:2022nmt}%
  \BibitemOpen
  \bibfield  {author} {\bibinfo {author} {\bibfnamefont {B.}~\bibnamefont
  {Zhou}}, \bibinfo {author} {\bibfnamefont {L.}~\bibnamefont {Reali}},
  \bibinfo {author} {\bibfnamefont {E.}~\bibnamefont {Berti}}, \bibinfo
  {author} {\bibfnamefont {M.}~\bibnamefont {\c{C}al\i{}\c{s}kan}}, \bibinfo
  {author} {\bibfnamefont {C.}~\bibnamefont {Creque-Sarbinowski}}, \bibinfo
  {author} {\bibfnamefont {M.}~\bibnamefont {Kamionkowski}}, \ and\ \bibinfo
  {author} {\bibfnamefont {B.~S.}\ \bibnamefont {Sathyaprakash}},\ }\href@noop
  {} {\  (\bibinfo {year} {2022}{\natexlab{a}})},\ \Eprint
  {http://arxiv.org/abs/2209.01310} {arXiv:2209.01310 [gr-qc]} \BibitemShut
  {NoStop}%
\bibitem [{\citenamefont {Zhou}\ \emph
  {et~al.}(2022{\natexlab{b}})\citenamefont {Zhou}, \citenamefont {Reali},
  \citenamefont {Berti}, \citenamefont {\c{C}al\i{}\c{s}kan}, \citenamefont
  {Creque-Sarbinowski}, \citenamefont {Kamionkowski},\ and\ \citenamefont
  {Sathyaprakash}}]{Zhou:2022otw}%
  \BibitemOpen
  \bibfield  {author} {\bibinfo {author} {\bibfnamefont {B.}~\bibnamefont
  {Zhou}}, \bibinfo {author} {\bibfnamefont {L.}~\bibnamefont {Reali}},
  \bibinfo {author} {\bibfnamefont {E.}~\bibnamefont {Berti}}, \bibinfo
  {author} {\bibfnamefont {M.}~\bibnamefont {\c{C}al\i{}\c{s}kan}}, \bibinfo
  {author} {\bibfnamefont {C.}~\bibnamefont {Creque-Sarbinowski}}, \bibinfo
  {author} {\bibfnamefont {M.}~\bibnamefont {Kamionkowski}}, \ and\ \bibinfo
  {author} {\bibfnamefont {B.~S.}\ \bibnamefont {Sathyaprakash}},\ }\href@noop
  {} {\  (\bibinfo {year} {2022}{\natexlab{b}})},\ \Eprint
  {http://arxiv.org/abs/2209.01221} {arXiv:2209.01221 [gr-qc]} \BibitemShut
  {NoStop}%
\bibitem [{\citenamefont {Reali}\ \emph {et~al.}(2022)\citenamefont {Reali},
  \citenamefont {Antonelli}, \citenamefont {Cotesta}, \citenamefont
  {Borhanian}, \citenamefont {\c{C}al\i{}\c{s}kan}, \citenamefont {Berti},\
  and\ \citenamefont {Sathyaprakash}}]{Reali:2022aps}%
  \BibitemOpen
  \bibfield  {author} {\bibinfo {author} {\bibfnamefont {L.}~\bibnamefont
  {Reali}}, \bibinfo {author} {\bibfnamefont {A.}~\bibnamefont {Antonelli}},
  \bibinfo {author} {\bibfnamefont {R.}~\bibnamefont {Cotesta}}, \bibinfo
  {author} {\bibfnamefont {S.}~\bibnamefont {Borhanian}}, \bibinfo {author}
  {\bibfnamefont {M.}~\bibnamefont {\c{C}al\i{}\c{s}kan}}, \bibinfo {author}
  {\bibfnamefont {E.}~\bibnamefont {Berti}}, \ and\ \bibinfo {author}
  {\bibfnamefont {B.~S.}\ \bibnamefont {Sathyaprakash}},\ }\href@noop {} {\
  (\bibinfo {year} {2022})},\ \Eprint {http://arxiv.org/abs/2209.13452}
  {arXiv:2209.13452 [gr-qc]} \BibitemShut {NoStop}%
\bibitem [{\citenamefont {Janquart}\ \emph {et~al.}(2022)\citenamefont
  {Janquart}, \citenamefont {Baka}, \citenamefont {Samajdar}, \citenamefont
  {Dietrich},\ and\ \citenamefont {Van Den~Broeck}}]{Janquart:2022nyz}%
  \BibitemOpen
  \bibfield  {author} {\bibinfo {author} {\bibfnamefont {J.}~\bibnamefont
  {Janquart}}, \bibinfo {author} {\bibfnamefont {T.}~\bibnamefont {Baka}},
  \bibinfo {author} {\bibfnamefont {A.}~\bibnamefont {Samajdar}}, \bibinfo
  {author} {\bibfnamefont {T.}~\bibnamefont {Dietrich}}, \ and\ \bibinfo
  {author} {\bibfnamefont {C.}~\bibnamefont {Van Den~Broeck}},\ }\href@noop {}
  {\  (\bibinfo {year} {2022})},\ \Eprint {http://arxiv.org/abs/2211.01304}
  {arXiv:2211.01304 [gr-qc]} \BibitemShut {NoStop}%
\bibitem [{\citenamefont {Zackay}\ \emph {et~al.}(2018)\citenamefont {Zackay},
  \citenamefont {Dai},\ and\ \citenamefont {Venumadhav}}]{Zackay:2018qdy}%
  \BibitemOpen
  \bibfield  {author} {\bibinfo {author} {\bibfnamefont {B.}~\bibnamefont
  {Zackay}}, \bibinfo {author} {\bibfnamefont {L.}~\bibnamefont {Dai}}, \ and\
  \bibinfo {author} {\bibfnamefont {T.}~\bibnamefont {Venumadhav}},\
  }\href@noop {} {\  (\bibinfo {year} {2018})},\ \Eprint
  {http://arxiv.org/abs/1806.08792} {arXiv:1806.08792 [astro-ph.IM]}
  \BibitemShut {NoStop}%
\bibitem [{\citenamefont {Dai}\ \emph {et~al.}(2018)\citenamefont {Dai},
  \citenamefont {Venumadhav},\ and\ \citenamefont {Zackay}}]{Dai:2018dca}%
  \BibitemOpen
  \bibfield  {author} {\bibinfo {author} {\bibfnamefont {L.}~\bibnamefont
  {Dai}}, \bibinfo {author} {\bibfnamefont {T.}~\bibnamefont {Venumadhav}}, \
  and\ \bibinfo {author} {\bibfnamefont {B.}~\bibnamefont {Zackay}},\
  }\href@noop {} {\  (\bibinfo {year} {2018})},\ \Eprint
  {http://arxiv.org/abs/1806.08793} {arXiv:1806.08793 [gr-qc]} \BibitemShut
  {NoStop}%
\bibitem [{\citenamefont {Leslie}\ \emph {et~al.}(2021)\citenamefont {Leslie},
  \citenamefont {Dai},\ and\ \citenamefont {Pratten}}]{Leslie:2021ssu}%
  \BibitemOpen
  \bibfield  {author} {\bibinfo {author} {\bibfnamefont {N.}~\bibnamefont
  {Leslie}}, \bibinfo {author} {\bibfnamefont {L.}~\bibnamefont {Dai}}, \ and\
  \bibinfo {author} {\bibfnamefont {G.}~\bibnamefont {Pratten}},\ }\href
  {\doibase 10.1103/PhysRevD.104.123030} {\bibfield  {journal} {\bibinfo
  {journal} {Phys. Rev. D}\ }\textbf {\bibinfo {volume} {104}},\ \bibinfo
  {pages} {123030} (\bibinfo {year} {2021})},\ \Eprint
  {http://arxiv.org/abs/2109.09872} {arXiv:2109.09872 [astro-ph.IM]}
  \BibitemShut {NoStop}%
\bibitem [{\citenamefont {Morisaki}(2021)}]{Morisaki:2021ngj}%
  \BibitemOpen
  \bibfield  {author} {\bibinfo {author} {\bibfnamefont {S.}~\bibnamefont
  {Morisaki}},\ }\href {\doibase 10.1103/PhysRevD.104.044062} {\bibfield
  {journal} {\bibinfo  {journal} {Phys. Rev. D}\ }\textbf {\bibinfo {volume}
  {104}},\ \bibinfo {pages} {044062} (\bibinfo {year} {2021})},\ \Eprint
  {http://arxiv.org/abs/2104.07813} {arXiv:2104.07813 [gr-qc]} \BibitemShut
  {NoStop}%
\bibitem [{\citenamefont {Gao}\ \emph {et~al.}(2022)\citenamefont {Gao},
  \citenamefont {Hayes}, \citenamefont {Croke}, \citenamefont {Messenger},\
  and\ \citenamefont {Veitch}}]{Gao:2021rxg}%
  \BibitemOpen
  \bibfield  {author} {\bibinfo {author} {\bibfnamefont {S.}~\bibnamefont
  {Gao}}, \bibinfo {author} {\bibfnamefont {F.}~\bibnamefont {Hayes}}, \bibinfo
  {author} {\bibfnamefont {S.}~\bibnamefont {Croke}}, \bibinfo {author}
  {\bibfnamefont {C.}~\bibnamefont {Messenger}}, \ and\ \bibinfo {author}
  {\bibfnamefont {J.}~\bibnamefont {Veitch}},\ }\href {\doibase
  10.1103/PhysRevResearch.4.023006} {\bibfield  {journal} {\bibinfo  {journal}
  {Phys. Rev. Res.}\ }\textbf {\bibinfo {volume} {4}},\ \bibinfo {pages}
  {023006} (\bibinfo {year} {2022})},\ \Eprint
  {http://arxiv.org/abs/2109.01535} {arXiv:2109.01535 [quant-ph]} \BibitemShut
  {NoStop}%
\bibitem [{\citenamefont {{Jimenez Rezende}}\ and\ \citenamefont
  {{Mohamed}}(2015)}]{2015arXiv150505770J}%
  \BibitemOpen
  \bibfield  {author} {\bibinfo {author} {\bibfnamefont {D.}~\bibnamefont
  {{Jimenez Rezende}}}\ and\ \bibinfo {author} {\bibfnamefont {S.}~\bibnamefont
  {{Mohamed}}},\ }\href@noop {} {\bibfield  {journal} {\bibinfo  {journal}
  {arXiv e-prints}\ ,\ \bibinfo {eid} {arXiv:1505.05770}} (\bibinfo {year}
  {2015})},\ \Eprint {http://arxiv.org/abs/1505.05770} {arXiv:1505.05770
  [stat.ML]} \BibitemShut {NoStop}%
\bibitem [{\citenamefont {{Kingma}}\ \emph {et~al.}(2016)\citenamefont
  {{Kingma}}, \citenamefont {{Salimans}}, \citenamefont {{Jozefowicz}},
  \citenamefont {{Chen}}, \citenamefont {{Sutskever}},\ and\ \citenamefont
  {{Welling}}}]{2016arXiv160604934K}%
  \BibitemOpen
  \bibfield  {author} {\bibinfo {author} {\bibfnamefont {D.~P.}\ \bibnamefont
  {{Kingma}}}, \bibinfo {author} {\bibfnamefont {T.}~\bibnamefont
  {{Salimans}}}, \bibinfo {author} {\bibfnamefont {R.}~\bibnamefont
  {{Jozefowicz}}}, \bibinfo {author} {\bibfnamefont {X.}~\bibnamefont
  {{Chen}}}, \bibinfo {author} {\bibfnamefont {I.}~\bibnamefont {{Sutskever}}},
  \ and\ \bibinfo {author} {\bibfnamefont {M.}~\bibnamefont {{Welling}}},\
  }\href@noop {} {\bibfield  {journal} {\bibinfo  {journal} {arXiv e-prints}\
  ,\ \bibinfo {eid} {arXiv:1606.04934}} (\bibinfo {year} {2016})},\ \Eprint
  {http://arxiv.org/abs/1606.04934} {arXiv:1606.04934 [cs.LG]} \BibitemShut
  {NoStop}%
\bibitem [{\citenamefont {{Papamakarios}}\ \emph {et~al.}(2017)\citenamefont
  {{Papamakarios}}, \citenamefont {{Pavlakou}},\ and\ \citenamefont
  {{Murray}}}]{2017arXiv170507057P}%
  \BibitemOpen
  \bibfield  {author} {\bibinfo {author} {\bibfnamefont {G.}~\bibnamefont
  {{Papamakarios}}}, \bibinfo {author} {\bibfnamefont {T.}~\bibnamefont
  {{Pavlakou}}}, \ and\ \bibinfo {author} {\bibfnamefont {I.}~\bibnamefont
  {{Murray}}},\ }\href@noop {} {\bibfield  {journal} {\bibinfo  {journal}
  {arXiv e-prints}\ ,\ \bibinfo {eid} {arXiv:1705.07057}} (\bibinfo {year}
  {2017})},\ \Eprint {http://arxiv.org/abs/1705.07057} {arXiv:1705.07057
  [stat.ML]} \BibitemShut {NoStop}%
\bibitem [{\citenamefont {Cuoco}\ \emph {et~al.}(2021)\citenamefont {Cuoco}
  \emph {et~al.}}]{Cuoco:2020ogp}%
  \BibitemOpen
  \bibfield  {author} {\bibinfo {author} {\bibfnamefont {E.}~\bibnamefont
  {Cuoco}} \emph {et~al.},\ }\href {\doibase 10.1088/2632-2153/abb93a}
  {\bibfield  {journal} {\bibinfo  {journal} {Mach. Learn. Sci. Tech.}\
  }\textbf {\bibinfo {volume} {2}},\ \bibinfo {pages} {011002} (\bibinfo {year}
  {2021})},\ \Eprint {http://arxiv.org/abs/2005.03745} {arXiv:2005.03745
  [astro-ph.HE]} \BibitemShut {NoStop}%
\bibitem [{\citenamefont {Cranmer}\ \emph {et~al.}(2020)\citenamefont
  {Cranmer}, \citenamefont {Brehmer},\ and\ \citenamefont
  {Louppe}}]{Cranmer:2020frontier}%
  \BibitemOpen
  \bibfield  {author} {\bibinfo {author} {\bibfnamefont {K.}~\bibnamefont
  {Cranmer}}, \bibinfo {author} {\bibfnamefont {J.}~\bibnamefont {Brehmer}}, \
  and\ \bibinfo {author} {\bibfnamefont {G.}~\bibnamefont {Louppe}},\
  }\href@noop {} {\bibfield  {journal} {\bibinfo  {journal} {Proceedings of the
  National Academy of Sciences}\ }\textbf {\bibinfo {volume} {117}},\ \bibinfo
  {pages} {30055} (\bibinfo {year} {2020})}\BibitemShut {NoStop}%
\bibitem [{\citenamefont {Delaunoy}\ \emph {et~al.}(2020)\citenamefont
  {Delaunoy}, \citenamefont {Wehenkel}, \citenamefont {Hinderer}, \citenamefont
  {Nissanke}, \citenamefont {Weniger}, \citenamefont {Williamson},\ and\
  \citenamefont {Louppe}}]{Delaunoy:2020zcu}%
  \BibitemOpen
  \bibfield  {author} {\bibinfo {author} {\bibfnamefont {A.}~\bibnamefont
  {Delaunoy}}, \bibinfo {author} {\bibfnamefont {A.}~\bibnamefont {Wehenkel}},
  \bibinfo {author} {\bibfnamefont {T.}~\bibnamefont {Hinderer}}, \bibinfo
  {author} {\bibfnamefont {S.}~\bibnamefont {Nissanke}}, \bibinfo {author}
  {\bibfnamefont {C.}~\bibnamefont {Weniger}}, \bibinfo {author} {\bibfnamefont
  {A.~R.}\ \bibnamefont {Williamson}}, \ and\ \bibinfo {author} {\bibfnamefont
  {G.}~\bibnamefont {Louppe}},\ }\href@noop {} {\  (\bibinfo {year} {2020})},\
  \Eprint {http://arxiv.org/abs/2010.12931} {arXiv:2010.12931 [astro-ph.IM]}
  \BibitemShut {NoStop}%
\bibitem [{\citenamefont {Green}\ \emph {et~al.}(2020)\citenamefont {Green},
  \citenamefont {Simpson},\ and\ \citenamefont {Gair}}]{Green:2020hst}%
  \BibitemOpen
  \bibfield  {author} {\bibinfo {author} {\bibfnamefont {S.~R.}\ \bibnamefont
  {Green}}, \bibinfo {author} {\bibfnamefont {C.}~\bibnamefont {Simpson}}, \
  and\ \bibinfo {author} {\bibfnamefont {J.}~\bibnamefont {Gair}},\ }\href
  {\doibase 10.1103/PhysRevD.102.104057} {\bibfield  {journal} {\bibinfo
  {journal} {Phys. Rev. D}\ }\textbf {\bibinfo {volume} {102}},\ \bibinfo
  {pages} {104057} (\bibinfo {year} {2020})},\ \Eprint
  {http://arxiv.org/abs/2002.07656} {arXiv:2002.07656 [astro-ph.IM]}
  \BibitemShut {NoStop}%
\bibitem [{\citenamefont {Green}\ and\ \citenamefont
  {Gair}(2021)}]{Green:2020dnx}%
  \BibitemOpen
  \bibfield  {author} {\bibinfo {author} {\bibfnamefont {S.~R.}\ \bibnamefont
  {Green}}\ and\ \bibinfo {author} {\bibfnamefont {J.}~\bibnamefont {Gair}},\
  }\href {\doibase 10.1088/2632-2153/abfaed} {\bibfield  {journal} {\bibinfo
  {journal} {Mach. Learn. Sci. Tech.}\ }\textbf {\bibinfo {volume} {2}},\
  \bibinfo {pages} {03LT01} (\bibinfo {year} {2021})},\ \Eprint
  {http://arxiv.org/abs/2008.03312} {arXiv:2008.03312 [astro-ph.IM]}
  \BibitemShut {NoStop}%
\bibitem [{\citenamefont {Dax}\ \emph {et~al.}(2021{\natexlab{a}})\citenamefont
  {Dax}, \citenamefont {Green}, \citenamefont {Gair}, \citenamefont {Deistler},
  \citenamefont {Sch\"olkopf},\ and\ \citenamefont {Macke}}]{Dax:2021myb}%
  \BibitemOpen
  \bibfield  {author} {\bibinfo {author} {\bibfnamefont {M.}~\bibnamefont
  {Dax}}, \bibinfo {author} {\bibfnamefont {S.~R.}\ \bibnamefont {Green}},
  \bibinfo {author} {\bibfnamefont {J.}~\bibnamefont {Gair}}, \bibinfo {author}
  {\bibfnamefont {M.}~\bibnamefont {Deistler}}, \bibinfo {author}
  {\bibfnamefont {B.}~\bibnamefont {Sch\"olkopf}}, \ and\ \bibinfo {author}
  {\bibfnamefont {J.~H.}\ \bibnamefont {Macke}},\ }\href@noop {} {\  (\bibinfo
  {year} {2021}{\natexlab{a}})},\ \Eprint {http://arxiv.org/abs/2111.13139}
  {arXiv:2111.13139 [cs.LG]} \BibitemShut {NoStop}%
\bibitem [{\citenamefont {Dax}\ \emph {et~al.}(2021{\natexlab{b}})\citenamefont
  {Dax}, \citenamefont {Green}, \citenamefont {Gair}, \citenamefont {Macke},
  \citenamefont {Buonanno},\ and\ \citenamefont {Sch\"olkopf}}]{Dax:2021tsq}%
  \BibitemOpen
  \bibfield  {author} {\bibinfo {author} {\bibfnamefont {M.}~\bibnamefont
  {Dax}}, \bibinfo {author} {\bibfnamefont {S.~R.}\ \bibnamefont {Green}},
  \bibinfo {author} {\bibfnamefont {J.}~\bibnamefont {Gair}}, \bibinfo {author}
  {\bibfnamefont {J.~H.}\ \bibnamefont {Macke}}, \bibinfo {author}
  {\bibfnamefont {A.}~\bibnamefont {Buonanno}}, \ and\ \bibinfo {author}
  {\bibfnamefont {B.}~\bibnamefont {Sch\"olkopf}},\ }\href {\doibase
  10.1103/PhysRevLett.127.241103} {\bibfield  {journal} {\bibinfo  {journal}
  {Phys. Rev. Lett.}\ }\textbf {\bibinfo {volume} {127}},\ \bibinfo {pages}
  {241103} (\bibinfo {year} {2021}{\natexlab{b}})},\ \Eprint
  {http://arxiv.org/abs/2106.12594} {arXiv:2106.12594 [gr-qc]} \BibitemShut
  {NoStop}%
\bibitem [{\citenamefont {Dax}\ \emph {et~al.}(2022)\citenamefont {Dax},
  \citenamefont {Green}, \citenamefont {Gair}, \citenamefont {P\"urrer},
  \citenamefont {Wildberger}, \citenamefont {Macke}, \citenamefont {Buonanno},\
  and\ \citenamefont {Sch\"olkopf}}]{Dax:2022pxd}%
  \BibitemOpen
  \bibfield  {author} {\bibinfo {author} {\bibfnamefont {M.}~\bibnamefont
  {Dax}}, \bibinfo {author} {\bibfnamefont {S.~R.}\ \bibnamefont {Green}},
  \bibinfo {author} {\bibfnamefont {J.}~\bibnamefont {Gair}}, \bibinfo {author}
  {\bibfnamefont {M.}~\bibnamefont {P\"urrer}}, \bibinfo {author}
  {\bibfnamefont {J.}~\bibnamefont {Wildberger}}, \bibinfo {author}
  {\bibfnamefont {J.~H.}\ \bibnamefont {Macke}}, \bibinfo {author}
  {\bibfnamefont {A.}~\bibnamefont {Buonanno}}, \ and\ \bibinfo {author}
  {\bibfnamefont {B.}~\bibnamefont {Sch\"olkopf}},\ }\href@noop {} {\
  (\bibinfo {year} {2022})},\ \Eprint {http://arxiv.org/abs/2210.05686}
  {arXiv:2210.05686 [gr-qc]} \BibitemShut {NoStop}%
\bibitem [{\citenamefont {Williams}\ \emph {et~al.}(2021)\citenamefont
  {Williams}, \citenamefont {Veitch},\ and\ \citenamefont
  {Messenger}}]{Williams:2021qyt}%
  \BibitemOpen
  \bibfield  {author} {\bibinfo {author} {\bibfnamefont {M.~J.}\ \bibnamefont
  {Williams}}, \bibinfo {author} {\bibfnamefont {J.}~\bibnamefont {Veitch}}, \
  and\ \bibinfo {author} {\bibfnamefont {C.}~\bibnamefont {Messenger}},\ }\href
  {\doibase 10.1103/PhysRevD.103.103006} {\bibfield  {journal} {\bibinfo
  {journal} {Phys. Rev. D}\ }\textbf {\bibinfo {volume} {103}},\ \bibinfo
  {pages} {103006} (\bibinfo {year} {2021})},\ \Eprint
  {http://arxiv.org/abs/2102.11056} {arXiv:2102.11056 [gr-qc]} \BibitemShut
  {NoStop}%
\bibitem [{\citenamefont {Kobyzev}\ \emph {et~al.}(2021)\citenamefont
  {Kobyzev}, \citenamefont {Prince},\ and\ \citenamefont {Brubaker}}]{9089305}%
  \BibitemOpen
  \bibfield  {author} {\bibinfo {author} {\bibfnamefont {I.}~\bibnamefont
  {Kobyzev}}, \bibinfo {author} {\bibfnamefont {S.~J.}\ \bibnamefont {Prince}},
  \ and\ \bibinfo {author} {\bibfnamefont {M.~A.}\ \bibnamefont {Brubaker}},\
  }\href {\doibase 10.1109/TPAMI.2020.2992934} {\bibfield  {journal} {\bibinfo
  {journal} {IEEE Transactions on Pattern Analysis and Machine Intelligence}\
  }\textbf {\bibinfo {volume} {43}},\ \bibinfo {pages} {3964} (\bibinfo {year}
  {2021})}\BibitemShut {NoStop}%
\bibitem [{\citenamefont {{Papamakarios}}\ \emph {et~al.}(2019)\citenamefont
  {{Papamakarios}}, \citenamefont {{Nalisnick}}, \citenamefont {{Jimenez
  Rezende}}, \citenamefont {{Mohamed}},\ and\ \citenamefont
  {{Lakshminarayanan}}}]{2019arXiv191202762P}%
  \BibitemOpen
  \bibfield  {author} {\bibinfo {author} {\bibfnamefont {G.}~\bibnamefont
  {{Papamakarios}}}, \bibinfo {author} {\bibfnamefont {E.}~\bibnamefont
  {{Nalisnick}}}, \bibinfo {author} {\bibfnamefont {D.}~\bibnamefont {{Jimenez
  Rezende}}}, \bibinfo {author} {\bibfnamefont {S.}~\bibnamefont {{Mohamed}}},
  \ and\ \bibinfo {author} {\bibfnamefont {B.}~\bibnamefont
  {{Lakshminarayanan}}},\ }\href@noop {} {\bibfield  {journal} {\bibinfo
  {journal} {arXiv e-prints}\ ,\ \bibinfo {eid} {arXiv:1912.02762}} (\bibinfo
  {year} {2019})},\ \Eprint {http://arxiv.org/abs/1912.02762} {arXiv:1912.02762
  [stat.ML]} \BibitemShut {NoStop}%
\bibitem [{\citenamefont {Chen}\ \emph {et~al.}(2018)\citenamefont {Chen},
  \citenamefont {Rubanova}, \citenamefont {Bettencourt},\ and\ \citenamefont
  {Duvenaud}}]{chen2018neural}%
  \BibitemOpen
  \bibfield  {author} {\bibinfo {author} {\bibfnamefont {R.~T.}\ \bibnamefont
  {Chen}}, \bibinfo {author} {\bibfnamefont {Y.}~\bibnamefont {Rubanova}},
  \bibinfo {author} {\bibfnamefont {J.}~\bibnamefont {Bettencourt}}, \ and\
  \bibinfo {author} {\bibfnamefont {D.~K.}\ \bibnamefont {Duvenaud}},\
  }\href@noop {} {\bibfield  {journal} {\bibinfo  {journal} {Advances in neural
  information processing systems}\ }\textbf {\bibinfo {volume} {31}} (\bibinfo
  {year} {2018})}\BibitemShut {NoStop}%
\bibitem [{\citenamefont {{Winkler}}\ \emph {et~al.}(2019)\citenamefont
  {{Winkler}}, \citenamefont {{Worrall}}, \citenamefont {{Hoogeboom}},\ and\
  \citenamefont {{Welling}}}]{2019arXiv191200042W}%
  \BibitemOpen
  \bibfield  {author} {\bibinfo {author} {\bibfnamefont {C.}~\bibnamefont
  {{Winkler}}}, \bibinfo {author} {\bibfnamefont {D.}~\bibnamefont
  {{Worrall}}}, \bibinfo {author} {\bibfnamefont {E.}~\bibnamefont
  {{Hoogeboom}}}, \ and\ \bibinfo {author} {\bibfnamefont {M.}~\bibnamefont
  {{Welling}}},\ }\href@noop {} {\bibfield  {journal} {\bibinfo  {journal}
  {arXiv e-prints}\ ,\ \bibinfo {eid} {arXiv:1912.00042}} (\bibinfo {year}
  {2019})},\ \Eprint {http://arxiv.org/abs/1912.00042} {arXiv:1912.00042
  [cs.LG]} \BibitemShut {NoStop}%
\bibitem [{\citenamefont {{Papamakarios}}\ and\ \citenamefont
  {{Murray}}(2016)}]{2016arXiv160506376P}%
  \BibitemOpen
  \bibfield  {author} {\bibinfo {author} {\bibfnamefont {G.}~\bibnamefont
  {{Papamakarios}}}\ and\ \bibinfo {author} {\bibfnamefont {I.}~\bibnamefont
  {{Murray}}},\ }\href@noop {} {\bibfield  {journal} {\bibinfo  {journal}
  {arXiv e-prints}\ ,\ \bibinfo {eid} {arXiv:1605.06376}} (\bibinfo {year}
  {2016})},\ \Eprint {http://arxiv.org/abs/1605.06376} {arXiv:1605.06376
  [stat.ML]} \BibitemShut {NoStop}%
\bibitem [{\citenamefont {Ha}\ \emph {et~al.}(2017)\citenamefont {Ha},
  \citenamefont {Dai},\ and\ \citenamefont {Le}}]{ha2017hypernetworks}%
  \BibitemOpen
  \bibfield  {author} {\bibinfo {author} {\bibfnamefont {D.}~\bibnamefont
  {Ha}}, \bibinfo {author} {\bibfnamefont {A.~M.}\ \bibnamefont {Dai}}, \ and\
  \bibinfo {author} {\bibfnamefont {Q.~V.}\ \bibnamefont {Le}},\ }in\ \href
  {https://openreview.net/forum?id=rkpACe1lx} {\emph {\bibinfo {booktitle}
  {International Conference on Learning Representations}}}\ (\bibinfo {year}
  {2017})\BibitemShut {NoStop}%
\bibitem [{\citenamefont {Zhuang}\ \emph {et~al.}(2021)\citenamefont {Zhuang},
  \citenamefont {Dvornek}, \citenamefont {sekhar tatikonda},\ and\
  \citenamefont {s~Duncan}}]{zhuang2021mali}%
  \BibitemOpen
  \bibfield  {author} {\bibinfo {author} {\bibfnamefont {J.}~\bibnamefont
  {Zhuang}}, \bibinfo {author} {\bibfnamefont {N.~C.}\ \bibnamefont {Dvornek}},
  \bibinfo {author} {\bibnamefont {sekhar tatikonda}}, \ and\ \bibinfo {author}
  {\bibfnamefont {J.}~\bibnamefont {s~Duncan}},\ }in\ \href
  {https://openreview.net/forum?id=blfSjHeFM_e} {\emph {\bibinfo {booktitle}
  {International Conference on Learning Representations}}}\ (\bibinfo {year}
  {2021})\BibitemShut {NoStop}%
\bibitem [{\citenamefont {{Halko}}\ \emph {et~al.}(2009)\citenamefont
  {{Halko}}, \citenamefont {{Martinsson}},\ and\ \citenamefont
  {{Tropp}}}]{2009arXiv0909.4061H}%
  \BibitemOpen
  \bibfield  {author} {\bibinfo {author} {\bibfnamefont {N.}~\bibnamefont
  {{Halko}}}, \bibinfo {author} {\bibfnamefont {P.-G.}\ \bibnamefont
  {{Martinsson}}}, \ and\ \bibinfo {author} {\bibfnamefont {J.~A.}\
  \bibnamefont {{Tropp}}},\ }\href@noop {} {\bibfield  {journal} {\bibinfo
  {journal} {arXiv e-prints}\ ,\ \bibinfo {eid} {arXiv:0909.4061}} (\bibinfo
  {year} {2009})},\ \Eprint {http://arxiv.org/abs/0909.4061} {arXiv:0909.4061
  [math.NA]} \BibitemShut {NoStop}%
\bibitem [{\citenamefont {{Barsotti}}\ \emph {et~al.}(2021)\citenamefont
  {{Barsotti}}, \citenamefont {{Fritschel}}, \citenamefont {{Evans}},\ and\
  \citenamefont {{Gras}}}]{aLIGOdesign}%
  \BibitemOpen
  \bibfield  {author} {\bibinfo {author} {\bibfnamefont {L.}~\bibnamefont
  {{Barsotti}}}, \bibinfo {author} {\bibfnamefont {P.}~\bibnamefont
  {{Fritschel}}}, \bibinfo {author} {\bibfnamefont {M.}~\bibnamefont
  {{Evans}}}, \ and\ \bibinfo {author} {\bibfnamefont {S.}~\bibnamefont
  {{Gras}}},\ }\href@noop {} {\enquote {\bibinfo {title} {{Advanced LIGO
  anticipated sensitivity curves}},}\ }\bibinfo {howpublished}
  {\url{https://dcc.ligo.org/LIGO-T1800044/public}} (\bibinfo {year}
  {2021})\BibitemShut {NoStop}%
\bibitem [{\citenamefont {Khan}\ \emph {et~al.}(2016)\citenamefont {Khan},
  \citenamefont {Husa}, \citenamefont {Hannam}, \citenamefont {Ohme},
  \citenamefont {P\"urrer}, \citenamefont {Jim\'enez~Forteza},\ and\
  \citenamefont {Boh\'e}}]{Khan:2015jqa}%
  \BibitemOpen
  \bibfield  {author} {\bibinfo {author} {\bibfnamefont {S.}~\bibnamefont
  {Khan}}, \bibinfo {author} {\bibfnamefont {S.}~\bibnamefont {Husa}}, \bibinfo
  {author} {\bibfnamefont {M.}~\bibnamefont {Hannam}}, \bibinfo {author}
  {\bibfnamefont {F.}~\bibnamefont {Ohme}}, \bibinfo {author} {\bibfnamefont
  {M.}~\bibnamefont {P\"urrer}}, \bibinfo {author} {\bibfnamefont
  {X.}~\bibnamefont {Jim\'enez~Forteza}}, \ and\ \bibinfo {author}
  {\bibfnamefont {A.}~\bibnamefont {Boh\'e}},\ }\href {\doibase
  10.1103/PhysRevD.93.044007} {\bibfield  {journal} {\bibinfo  {journal} {Phys.
  Rev. D}\ }\textbf {\bibinfo {volume} {93}},\ \bibinfo {pages} {044007}
  (\bibinfo {year} {2016})},\ \Eprint {http://arxiv.org/abs/1508.07253}
  {arXiv:1508.07253 [gr-qc]} \BibitemShut {NoStop}%
\bibitem [{\citenamefont {Kolmus}\ \emph {et~al.}(2022)\citenamefont {Kolmus},
  \citenamefont {Baltus}, \citenamefont {Janquart}, \citenamefont {van
  Laarhoven}, \citenamefont {Caudill},\ and\ \citenamefont
  {Heskes}}]{Kolmus:2021buf}%
  \BibitemOpen
  \bibfield  {author} {\bibinfo {author} {\bibfnamefont {A.}~\bibnamefont
  {Kolmus}}, \bibinfo {author} {\bibfnamefont {G.}~\bibnamefont {Baltus}},
  \bibinfo {author} {\bibfnamefont {J.}~\bibnamefont {Janquart}}, \bibinfo
  {author} {\bibfnamefont {T.}~\bibnamefont {van Laarhoven}}, \bibinfo {author}
  {\bibfnamefont {S.}~\bibnamefont {Caudill}}, \ and\ \bibinfo {author}
  {\bibfnamefont {T.}~\bibnamefont {Heskes}},\ }\href {\doibase
  10.1103/PhysRevD.106.023032} {\bibfield  {journal} {\bibinfo  {journal}
  {Phys. Rev. D}\ }\textbf {\bibinfo {volume} {106}},\ \bibinfo {pages}
  {023032} (\bibinfo {year} {2022})},\ \Eprint
  {http://arxiv.org/abs/2111.00833} {arXiv:2111.00833 [gr-qc]} \BibitemShut
  {NoStop}%
\end{thebibliography}%
\end{document}